\documentclass{jaa}
\usepackage{graphicx, url, hyperref, natbib, subcaption}
\usepackage{caption}
\usepackage{booktabs}
\usepackage{xcolor}
\usepackage[figuresright]{rotating}

\begin{document}

\title{A Multi-dimensional Analysis of the Open Clusters NGC 2423 and NGC 2482 with Gaia DR3}

\author{
W. H. Elsanhoury\textsuperscript{1,*},
D. C. Çınar\textsuperscript{2},
A. Ahmed\textsuperscript{3}
}

\affilOne{\textsuperscript{1}
Department of Astronomy, National Research Institute of Astronomy \& Geophysics (NRIAG), 11421, Helwan, Cairo, Egypt.\\}

\affilTwo{\textsuperscript{2}
Programme of Astronomy and Space Sciences, Institute of Graduate Studies in Science, Istanbul University, Beyazıt, Istanbul 34116, Türkiye.\\}

\affilThree{\textsuperscript{3}
Department of Astronomy, Faculty of Science, Cairo University, Giza 12613, Egypt.}

\twocolumn[{

\maketitle

\corres{elsanhoury@nriag.sci.eg}


\begin{abstract}
We present a homogeneous photometric, astrometric, and kinematic study of the Hyades-like open clusters NGC 2423 and NGC 2482 based on {\it Gaia} DR3 data. Cluster memberships are determined using the UPMASK algorithm in the multi-dimensional parameter space of $(\alpha,\delta,\varpi,\mu_{\alpha}\cos\delta,\mu_{\delta})$, with radial velocities incorporated when available. Gaussian modelling of the astrometric distributions yields mean trigonometric parallaxes of $\varpi = 1.060 \pm 0.035$ mas and $0.735 \pm 0.061$ mas, corresponding to astrometric distances of $944 \pm 31$ pc and $1361 \pm 113$ pc for NGC 2423 and NGC 2482, respectively. Fundamental parameters are derived through PARSEC isochrone fitting, resulting in ages of $1250 \pm 100$ Myr and $445 \pm 50$ Myr, observed distance moduli $(m-M)_{\rm obs}=9.90 \pm 0.10$ and $10.80 \pm 0.20$ mag, and reddening values $E(G_{\rm BP}-G_{\rm RP})=0.07 \pm 0.01$ and $0.10 \pm 0.02$ mag, equivalent to $E(B-V)=0.054$ and $0.078$. The corresponding photometric distances are $892 \pm 50$ pc and $1311 \pm 146$ pc. We further investigate the internal and global kinematics of both clusters using velocity-ellipsoid parameters, the convergent-point method, and action--angular-momentum diagnostics. \textcolor{black}{The derived kinematic properties show that NGC 2482 closely matches the velocity-space characteristics of the Hyades stream, while NGC 2423 shares the characteristic rotational velocity and angular momentum of the Hyades-like population but is offset in its $U$ component. Both clusters occupy the angular-momentum range associated in the literature with Hyades-like clusters and bar-related resonances, although this association is based on their present-day kinematics and action-space locations rather than a direct dynamical demonstration of resonance trapping.
}
\end{abstract}

}]

\doinum{12.3456/s78910-011-012-3}
\artcitid{\#\#\#\#}
\volnum{000}
\year{0000}
\pgrange{1--}
\setcounter{page}{1}
\lp{1}

\section{Introduction}\label{Introduction}

The majority of Milky Way (MW) field stars are now thought to originate in star clusters (e.g., \citep{2003ARA&A..41...57L, 2013MNRAS.428.1950H}). After their formation, clusters gradually dissolve and release their member stars into the Galactic field at rates that depend on both their internal properties and their Galactic environment (e.g., Galactocentric distance). Early studies by \cite{1938ZaTsA..22...19A} and \cite{1940MNRAS.100..396S} suggested that clusters may leave behind small stellar remnants rather than evaporating completely.

Open clusters (OCs) form a population whose study is crucial for understanding the structure and evolution of the Galactic disk. In particular, old OCs provide insight into star formation in the early disk. Young OCs, especially those associated with H\,\textsc{ii} regions or with ages of order $10^7$ yr or less, have been widely used to trace spiral structure, whereas intermediate-age OCs are effective discriminants between standard stellar evolution models and models that include convective overshooting (see, e.g., \cite{1988A&A...193..148M, 1990A&A...240..262A, 1994A&AS..107...39C}).

Hyades-like open clusters are well represented in the Solar neighbourhood and provide a valuable benchmark for comparison with observations. In the literature, the terms ``Hyades moving group'' or ``Hyades-like clusters'' do not necessarily refer to coeval stellar populations or clusters sharing a genetic origin from the same molecular cloud. Instead, the classification denotes stellar groups that exhibit a coherent kinematic stream in both velocity and action space, primarily driven by Galactic resonances. 

These OCs typically display a rotational lag characterized by a $V$ space velocity component of $V \sim -20$ km s$^{-1}$ \citep{Dehnen1998, Famaey2005}, and their vertical angular momentum ($L_z$) is tightly clusters within a specific range, typically between $1700$ and $1900$~kpc~km~s$^{-1}$ \citep{Boffin2022}. Several well-known open clusters satisfy these stringent kinematic criteria and are thus classified as Hyades-like, including the Hyades itself, Praesepe, Coma Berenices, and NGC~752 \citep{Famaey2005, Boffin2022}. Despite spanning a wide range of ages, from $\sim 600$~Myr for the Hyades to $\sim 1.5$~Gyr for NGC~752, these clusters belong to the same kinematic family. This velocity-space coherence is widely attributed to the dynamical influence of the 2:1 inner Lindblad resonance within the Galactic disk.

Among the population of clusters proposed to share this kinematic signature, NGC 2423 and NGC 2482 were selected for the present analysis for two reasons. First, both were explicitly classified as Hyades-like on chemical and kinematic grounds by \citet{1979ApJ...234..168E}, prior to the availability of high-precision {\it Gaia} astrometry, but neither has since been the subject of a dedicated, homogeneous re-analysis of its membership, structure, and kinematics using {\it Gaia} DR3 -- unlike better-studied members of the group such as Praesepe, Coma Berenices, and NGC~752. Second, their ages ($\sim1.25$~Gyr and $\sim0.45$~Gyr, respectively; see Sections~1.1 and~1.2 below) place them at opposite ends of the age range spanned by previously confirmed Hyades-like clusters, making them a useful joint test of whether the shared angular-momentum properties of Hyades-like clusters holds independently of cluster age, as expected if the underlying mechanism is a Galactic resonance rather than a common formation event (this is examined quantitatively in the Discussion, Section~\ref{Introduction} onward).

\subsection{NGC 2423}
The Hyades and Praesepe clusters, the two main stellar concentrations associated with the Hyades group, satisfy both the kinematical and chemical criteria for group membership (e.g., \citealt{1970VA.....12..367E, 1977ApJ...215..812E}) and are comparatively rich in stars. In this context, NGC 2423 and NGC 2482 have been proposed as Hyades-like clusters. We follow \citet{1979ApJ...234..168E} in classifying NGC 2423 (also known as Collinder~153, MWSC~1283, Melotte~70, OCL~592, and Theia~1176) as Hyades-like.

NGC 2423 is a detached Galactic OC with a substantial population of relatively luminous members distributed over a wide area \citep{1976A&AS...26...13H}. It lies in the constellation Puppis at ($\alpha_{\rm J2000} = 07^{\rm h}\,37^{\rm m}\,16^{\rm s}.179$, $\delta_{\rm J2000} = -13^{\circ}\,51^{\prime}\,47^{\prime\prime}.738$), corresponding to Galactic coordinates ($l=230^{\circ}.496$, $b=3^{\circ}.577$). Owing to its proximity to the celestial equator, NGC 2423 is observable from both hemispheres during suitable seasons. Its integrated apparent magnitude is $V\approx 6.7$.

More recently, \citet{2024ApJ...968...22B} estimated an age of $\sim 957$~Myr, a metallicity of ${\rm [Fe/H]} = 0.18$~dex, and an extinction of $A_V=0.15$~mag for NGC 2423, and derived a distance of 860~pc (corresponding to $(m-M)_0\sim 9.7$~mag) from isochrone fitting to the blue edge of the main-sequence turn-off region. These results are broadly consistent with \citet{2021MNRAS.504..356D}, who found an age of $\log~(t/{\rm yr}) = 9.0371$, a distance of 924~pc, and ${\rm [Fe/H]}$ = 0.122. The mean radial velocity is $V_r\approx 20\,{\rm km\,s^{-1}}$ \citep{2024ApJ...968...22B}.

\subsection{NGC 2482}
NGC 2482 is a rich, faint OC in the fourth Galactic quadrant, located slightly above the Galactic plane at ($\alpha_{\rm J2000} = 07^{\rm h}\,55^{\rm m}\,14^{\rm s}.310$, $\delta_{\rm J2000} = -24^{\circ}\,16^{\prime}\,47^{\prime\prime}.642$), corresponding to ($l=242^{\circ}.015$, $b=2^{\circ}.374$). It is also catalogued as Collinder~166, ESO~494--3, MWSC~1382, OCL~653, and Theia~2811. NGC 2482 is a relatively sparse system and is slightly older than the Hyades ($\sim 600$~Myr), with age estimates in the literature ranging between $\sim 400$ and $\sim 800$~Myr. 

\textcolor{black}{Unlike NGC 2423, for which a recent, dedicated Gaia~DR3-based isochrone analysis exists (\citealt{2024ApJ...968...22B}), thefundamental parameters of NGC 2482 available in the literature are more heterogeneous in origin: \citet{2021MNRAS.504..356D} performed isochrone fitting, but using Gaia~DR2 astrometry and photometry, while the estimates listed under \citet{Hunt2024} and \citet{CantatGaudin2020}) rely on non-isochrone, machine-learning-based classification methods rather than direct CMD fitting. To our knowledge, no dedicated isochrone analysis of NGC 2482 based on Gaia~DR3 photometry and astrometry has yet been published; large-scale Gaia~DR3 all-sky catalogues (e.g., \citealt{Hunt2024}) provide only homogeneous, catalogue-level astrometric and non-isochrone-based parameters at the cluster level. This gap, specifically the absence of a dedicated Gaia~DR3 isochrone-based analysis for NGC 2482, is part of the motivation for the present re-analysis, carried out jointly and homogeneously with NGC 2423.}

Photoelectric $UBV$ photometry for 41 stars in the cluster field was obtained by \citet{1975A&AS...20...85M}, who derived a mean reddening of $E(B-V) = 0.04 \pm 0.02$ and an apparent distance modulus of $(m-M)$ = 9.50, corresponding to $d\approx 750$ pc.

\begin{table*}[!ht]
\centering
\renewcommand{\arraystretch}{1.1}
\caption{Comparison of literature parameters for the investigated clusters. Listed are equatorial coordinates ($\alpha$,~$\delta$; sexagesimal), cluster radius ($r$; arcmin), logarithmic age ($\log~(t/{\rm yr})$), distance ($d$; pc), colour excess ($E(B-V)$; mag), trigonometric parallax ($\varpi$; mas), proper motion components ($\mu_\alpha \cos\delta$,~$\mu_\delta$; mas\,yr$^{-1}$), number of members ($N$), metallicity ([Fe/H]; dex), and references.}
\label{tab:literature}
\setlength\tabcolsep{2pt}
\
\scalebox{0.85}{
\begin{tabular}{cccccccccccc}
\hline
$\alpha$ & $\delta$ & $r$ & $\log~(t)$ & $d$ & $E(B-V)$ & $\varpi$ & $\mu_\alpha\cos\delta$ & $\mu_\delta$ & $N$ & [Fe/H] & Ref. \\
(hh:mm:ss) & (dd:mm:ss) & (arcmin) & (yr) & (pc) & (mag) & (mas) & (mas\,yr$^{-1}$) & (mas\,yr$^{-1}$) &  & (dex) &  \\
\hline\hline
\multicolumn{12}{c}{NGC 2423} \\
\hline
07:37:17.80 & -13:52:19.0 & 10.55 & 9.10 & 944 & 0.054 & 1.060 $\pm$ 0.035 & -0.750 $\pm$ 0.134 & -3.578 $\pm$ 0.143 & 488 & 0.00 & 0 \\
07:37:16.18 & $-$13:51:47.8 & 41.42 & 9.29 & 910 & -- & $1.059\pm0.038$ & $-0.747\pm0.109$ & $-3.582\pm0.108$ & 571 & -- & 1 \\
07:37:11.76 & $-$13:51:46.8 & 16.08$^*$ & 9.04 & 943 & -- & $1.059\pm0.031$ & $-0.747\pm0.114$ & $-3.586\pm0.113$ & 398 & -- & 2 \\
07:37:10.25 & $-$13:50:03.8 & 16.80$^*$ & 9.04 & 924 & -- & $1.043\pm0.054$ & $-0.729\pm0.166$ & $-3.642\pm0.171$ & 425 & \textcolor{black}{0.122$\pm$0.038} & 3 \\
07:37:11.76 & $-$13:51:46.8 & 16.08$^*$ & -- & 933 & -- & $1.045\pm0.051$ & $-0.735\pm0.124$ & $-3.632\pm0.149$ & 430 & -- & 4 \\
07:37:11.76 & $-$13:51:46.8 & 16.08$^*$ & 9.04 & 943 & -- & $1.045\pm0.051$ & $-0.735\pm0.124$ & $-3.632\pm0.149$ & 406 & -- & 5 \\
07:37:06.00 & $-$13:52:18.0 & 8.00 & 8.87 & 766 & 0.10 & -- & $-1.720\pm4.870$ & $-0.510\pm6.080$ & 455 & -- & 6 \\
07:37:06.00 & $-$13:52:18.0 & 7.00 & -- & -- & -- & -- & $-1.340\pm2.140$ & $-0.910\pm3.710$ & 417 & -- & 7 \\
07:37:06.48 & $-$13:51:00.0 & 27.00 & 9.03 & 824 & 0.06 & -- & $-0.600$ & $-2.890$ & -- & 0.068 & 8 \\
\hline\hline
\multicolumn{12}{c}{NGC 2482} \\
\hline
07:55:07.10 & -24:17:43.3 & 40.70 & 8.65 & 1361 & 0.078 & 0.735 $\pm$ 0.061 & -4.669 $\pm$ 0.070 & 2.250 $\pm$ 0.065 & 116 & 0.00 & 0 \\
07:55:14.30 & $-$24:16:47.6 & 46.34 & 8.97 & 1290 & -- & $0.726\pm0.055$ &\textcolor{black}{-4.667$\pm$0.105} & $2.260\pm0.121$ & 192 & -- & 1 \\
07:55:08.88 & $-$24:15:46.8 & 8.94$^*$ & 8.54 & 1285 & -- & $0.734\pm0.040$ & $-4.667\pm0.050$ & $2.253\pm0.070$ & 113 & -- & 2 \\
07:55:10.80 & $-$24:15:45.7 & 8.94$^*$ & 8.67 & 1308 & -- & $0.727\pm0.061$ & $-4.708\pm0.118$ & $2.209\pm0.181$ & 126 & $0.031\pm0.139$ & 3 \\
07:55:08.88 & $-$24:15:46.8 & 8.94$^*$ & -- & 1340 & -- & $0.720\pm0.053$ & $-4.722\pm0.094$ & $2.189\pm0.100$ & 128 & -- & 4 \\
07:55:08.88 & $-$24:15:46.8 & 8.94$^*$ & 8.54 & 1285 & -- & $0.720\pm0.053$ & $-4.722\pm0.094$ & $2.189\pm0.100$ & 115 & -- & 5 \\
07:55:12.00 & $-$24:15:30.0 & 5.50 & 8.60 & 1343 & 0.09 & -- & $-3.610\pm4.830$ & $2.860\pm4.850$ & 210 & -- & 6 \\
07:55:12.00 & $-$24:15:30.0 & 6.00 & -- & -- & -- & -- & $-3.290\pm2.760$ & $3.290\pm2.450$ & 250 & -- & 7 \\
07:55:10.08 & $-$24:16:30.0 & 12.60 & 8.51 & 1344 & 0.09 & -- & $-5.100$ & $1.800$ & -- & $-0.081$ & 8 \\
$-$& $-$ & $-$& $-$& $-$& $-$& $-$& $-$& $-$& $-$& \textcolor{black}{$-$0.081$\pm$0.141} & 9\\
$-$ & $-$ & $-$& \textcolor{black}{8.61} & \textcolor{black}{1272} & \textcolor{black}{0.09} & $-$& $-$& $-$& $-$& \textcolor{black}{0.09} & 10\\
$-$& $-$ & $-$& $-$& $-$& $-$& $-$& $-$& $-$& $-$&  \textcolor{black}{$-$0.07$\pm$0.04} & 11\\
\hline
\end{tabular}}
\\
\noindent{\footnotesize
Ref.: \textcolor{black}{(0) This study;} (1) \cite{Hunt2024}; (2) \cite{Poggio_2021}; (3) \cite{Dias_2021}; (4) \cite{Cantat-Gaudin-Anders_2020}; (5) \cite{CantatGaudin2020}; (6) \cite{sampedro2017}; (7) \cite{Dias2014}; (8) \cite{kharchenko2013vizier};(9) \cite{2014A&A...562A..54C}; (10) \cite{2015PASP..127...31K} ; (11) \cite{2013MNRAS.431.3338R}; \\
$^*$ Radii enclosing 50\% of the cluster members.}
\end{table*}

\citet{1983JApA....4..117C} identified three probable giants as photometric members and estimated $E(B-V) = 0.09$ from the hotter stars and 0.11 from DDO (David Dunlap Observatory) photometry of the giants, both higher than the original estimate of 0.04 by \citet{1975A&AS...20...85M}. Using a larger sample of A-type stars, \citet{1997AJ....114.2556T} adopted $E(B-V) = 0.09$ for the giants and 0.10 for the main sequence. Because the colour--magnitude diagrams (CMD) of \citet{1975A&AS...20...85M} is poorly defined, a reliable distance estimate is difficult; using the DDO modulus (scaled to the Hyades distance) yields a highly uncertain $(m-M) = 10.3$.

Recent studies generally suggest that NGC 2482 is an intermediate-age cluster, with age estimates typically clustering around $\sim 400$~Myr \citep{2013MNRAS.431.3338R, 2015PASP..127...31K, 2014A&A...562A..54C}, and listed in Table~\ref{tab:literature}.

\textcolor{black}{The structure of this paper is organized as follows. In Section~2, we describe the observational data and source selection criteria obtained from the Gaia DR3 archive. Section~3 details the methodology for determining the cluster centers and assigning membership using the UPMASK algorithm. The structural properties, including radial density profiles and King-model fits, are examined in Section~4. Section~5 presents the astrometric properties of the clusters, while Section~6 derives their fundamental astrophysical parameters through isochrone fitting. The kinematic and dynamical analysis, including radial velocities and Galactic orbits, is presented in Section~7. Section~8 investigates the luminosity and mass functions, with particular emphasis on mass segregation. Section~9 discusses the kinematic properties of the clusters in the context of the Hyades moving group and Galactic resonance. Finally, the main results are summarized in Section~10.}
\section{Data}

This study is based on astrometric, photometric, and spectroscopic data from the {\it Gaia} DR3 archive \citep{GaiaDR3}. The photometric analysis uses the $G$, $G_{\rm BP}$, and $G_{\rm RP}$ bands. The astrometric parameters include equatorial coordinates ($\alpha,~\delta$), trigonometric parallaxes ($\varpi$), and proper-motion components ($\mu_{\alpha}\cos\delta,~\mu_{\delta}$). Radial velocities ($V_{\rm R}$) were included when available to improve the kinematic analysis, but their absence for a given star did not exclude it from the astrometric membership analysis: membership was assigned in the five-dimensional space $(\alpha,~\delta,~\varpi,~\mu_\alpha\cos\delta,~\mu_\delta)$, and radial velocities were required only for the subset of members entering the radial-velocity, three-dimensional kinematic, and orbital analyses of Section~\ref{sec:mean_rv} onward. All quantities were used together with their reported uncertainties. 

The initial source extraction was performed within circular regions of radius $40^{\prime}$ for NGC 2423 and $60^{\prime}$ for NGC 2482, centered on the nominal coordinates of \citet{Hunt2024}. We adopted cluster centers of $\alpha = 07^{\text{h}} 37^{\text{m}} 16^{\text{s}}$, $\delta = -13^{\circ} 51^{\prime} 47^{\prime\prime}$ for NGC 2423 and $\alpha = 07^{\text{h}} 55^{\text{m}} 14^{\text{s}}$, $\delta = -24^{\circ} 16^{\prime} 47^{\prime\prime}$ for NGC 2482. Following the {\it Gaia} DR3 recommendations for astrometrically well-behaved solutions, the query was restricted to sources with \texttt{astrometric\_params\_solved} $\in\{31,95\}$ (i.e., full five- or six-parameter astrometric solutions) and with a reported parallax, so that basic-model or incomplete astrometric solutions did not enter the analysis at the extraction stage. This selection yielded 87,227 sources toward NGC 2423 and 128,381 sources toward NGC 2482. Restricting this sample to the adopted limiting magnitude of $G \le 20.5$~mag (Section~\ref{sec:membership}) reduces these totals to 74,478 and 106,150 sources, respectively; for reference, an additional RUWE~$\le 1.4$ cut applied at this stage would further reduce the full $G \le 20.5$~mag sample to 71,933 and 101,996 sources. In the analysis as actually performed, however, this astrometric quality cut is not applied to the full, field-dominated sample, but only after membership assignment, to the probable cluster members alone (see Section~\ref{sec:membership}).

To characterize the photometric quality of the full extracted samples (i.e., before membership selection, and thus dominated in number by field stars) as a function of apparent magnitude, we binned all sources within each search radius by $G$ magnitude and computed, in each bin, the arithmetic mean of the individual per-star $G$ and $(G_{\rm BP}-G_{\rm RP})$ uncertainties reported by the {\it Gaia} DR3 archive. This is intended purely as a diagnostic of the magnitude-dependent photometric precision of the input catalogue and is used to justify the adopted limiting magnitude of $G\leq20.5$~mag for the membership analysis (Section~\ref{sec:membership}) and to indicate where photometric noise, rather than intrinsic scatter, dominates the faint end of the observed CMDs (Figure~\ref{CMDs}). The results are summarized in Table~\ref{tab:photometric_errors} as a function of $G$-band magnitude. 

The photometry is highly precise at the bright end ($G<18$ mag), where the mean $G$-band uncertainties ($\sigma_{\rm G}$) remain close to 0.003 mag for both clusters, consistent with the nominal {\it Gaia} DR3 performance for unsaturated sources. As expected, the uncertainties increase nonlinearly toward fainter magnitudes. For $G=20$--21 mag, $\sigma_{\rm G}$ reaches 0.0078--0.0102 mag, while the colour uncertainty ($\sigma_{G_{\rm BP}-G_{\rm RP}}$) increases to $\sim 0.16$--0.21 mag. Beyond $G=21$ mag, the photometric quality degrades further, with colour uncertainties exceeding 0.4 mag, reflecting the survey's detection limits and the increasing influence of background noise. \textcolor{black}{We emphasize that Table~\ref{tab:photometric_errors} characterizes the full extracted sample described above, down to the faintest sources retrieved from the archive ($G \le 23$~mag), and not the sample used in the membership analysis. Beyond the $G \le 20.5$~mag selection itself, no additional photometric quality cut (e.g., on $\sigma_{\rm G}$ or $\sigma_{G_{\rm BP}-G_{\rm RP}}$) was applied to the input sample prior to running UPMASK.}

\begin{table*}[!ht]
\centering
\footnotesize
\renewcommand{\arraystretch}{1.1}
\caption{\textcolor{black}{Mean photometric uncertainties in {\it Gaia} apparent magnitudes ($G$) and colours ($G_{\rm BP}-G_{\rm RP}$) for the full extracted sample toward NGC 2423 and NGC 2482.}}
\label{tab:photometric_errors}
\begin{tabular}{c|ccc|ccc}
\toprule
 & \multicolumn{3}{c|}{\textcolor{black}{NGC 2423}} & \multicolumn{3}{c}{\textcolor{black}{NGC 2482}} \\\cmidrule(lr){2-4} \cmidrule(lr){5-7}$G$ (mag) & $N$ & $\sigma_{{\rm G}}$ & $\sigma_{{G_{{\rm BP}}-G_{{\rm RP}}}}$ & $N$ & $\sigma_{{\rm G}}$ & $\sigma_{{G_{{\rm BP}}-G_{{\rm RP}}}}$ \\ \hline \hline
6--14 & 1854 & 0.0029 & 0.0054 & 2258 & 0.0028 & 0.0054 \\ 
14--15 & 1901 & 0.0029 & 0.0058 & 2645 & 0.0028 & 0.0055 \\ 
15--16 & 3742 & 0.0029 & 0.0064 & 5089 & 0.0029 & 0.0067 \\ 
16--17 & 6617 & 0.0029 & 0.0090 & 9363 & 0.0029 & 0.0081 \\ 
17--18 & 10766 & 0.0031 & 0.0168 & 15096 & 0.0030 & 0.0138 \\ 
18--19 & 15412 & 0.0035 & 0.0382 & 22041 & 0.0032 & 0.0299 \\ 
19--20 & 20038 & 0.0047 & 0.0822 & 29159 & 0.0040 & 0.0624 \\ 
20--21 & 24939 & 0.0102 & 0.2065 & 35960 & 0.0078 & 0.1615 \\ 
21--23 & 1958 & 0.0253 & 0.4202 & 6770 & 0.0232 & 0.4030 \\ 
\hline
Total/Error & 87227 & 0.0060 & 0.0974 & 128381 & 0.0057 & 0.0885 \\ \hline
\bottomrule
\end{tabular}
\end{table*}

\section{Cluster Membership Determination}\label{sec:membership}

Accurate determination of the astrophysical parameters of NGC 2423 and NGC 2482 requires a reliable identification of their stellar members, particularly in the presence of strong field-star contamination in the Galactic plane. Since members of an OC are coeval and share common kinematic properties, we utilized their proper motions and trigonometric parallaxes as the basis for membership assignment. To compute the membership probabilities ($P$), we employed the Unsupervised Photometric Membership Assignment in Stellar Clusters (\textsc{UPMASK}) algorithm \citep{Krone-Martins_2014}. This method identifies spatial and kinematic coherence in a multi-dimensional parameter space using $k$-means clustering and statistical resampling, without requiring prior assumptions about the cluster structure.

\begin{figure}[ht!]
    \centering
    \includegraphics[width=1\linewidth]{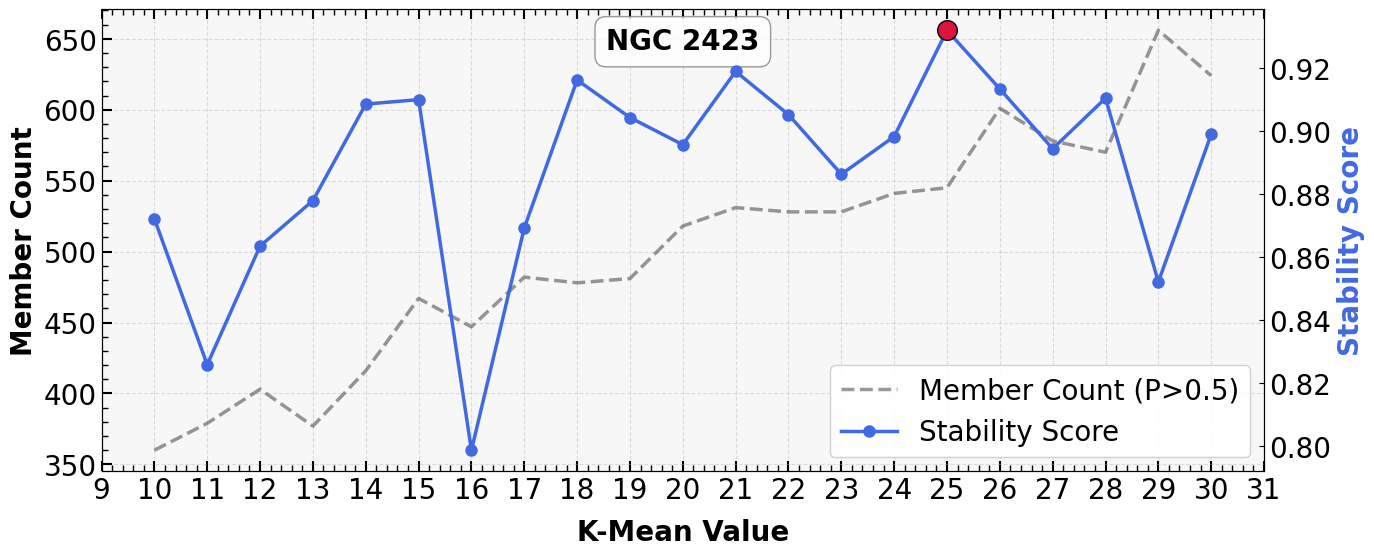}\\
    \includegraphics[width=1\linewidth]{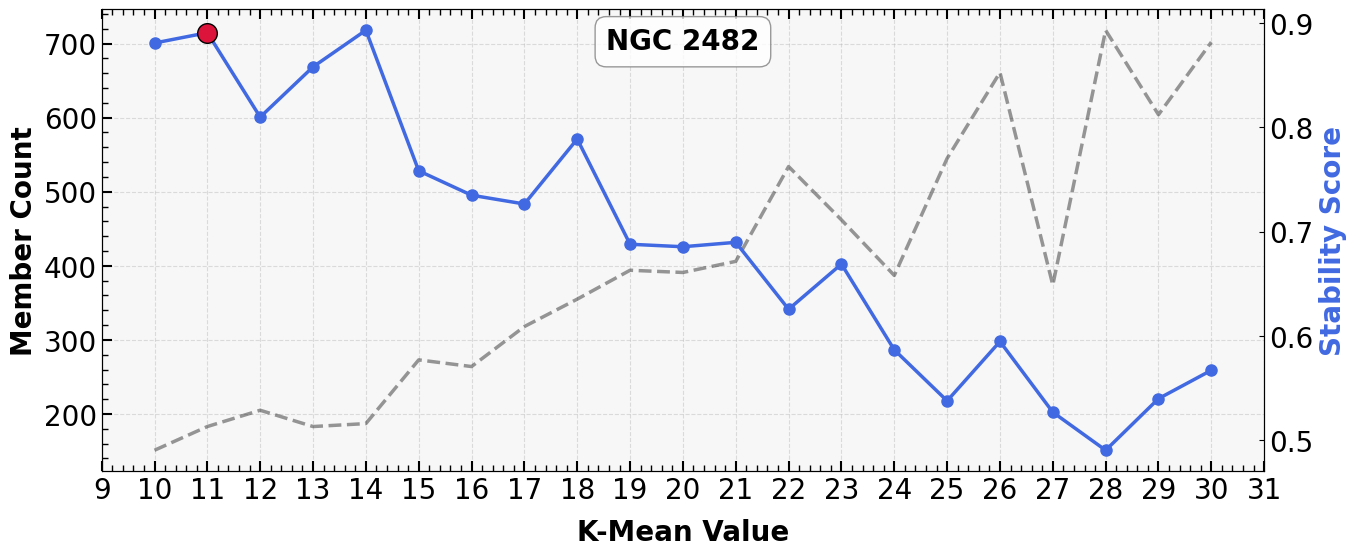}
    \caption{\textsc{UPMASK} stability analysis for NGC 2423 and NGC 2482 over the range $10 \leq k \leq 30$. Solid blue lines show the stability score, defined as the ratio of stars with $P > 0.8$ to those with $P > 0.5$. Dashed grey lines indicate the number of probable members ($P > 0.5$). Red circles mark the adopted values of $k$ (25 for NGC 2423 and 11 for NGC 2482).}

\label{fig:upmask_stability}
\end{figure}

The \textsc{UPMASK} analysis was executed using \textit{Gaia} DR3 astrometric data, specifically equatorial coordinates ($\alpha,~\delta$), proper-motion components ($\mu_{\alpha} \cos \delta,~\mu_{\delta}$), and trigonometric parallaxes ($\varpi$), including their respective uncertainties. \textsc{UPMASK} was run independently for each integer value of $k$ in the range $10 \leq k \leq 30$; for each $k$, the algorithm was repeated over 50 resampling iterations, and the resulting membership probability $P$ for every star was averaged over these iterations. For each $k$ we then computed a stability score, defined as the ratio of the number of stars with $P>0.8$ to the number of stars with $P>0.5$ -- i.e., how sharply the probability distribution separates into a high confidence core versus a looser moderate-probability sample. A value of this ratio close to unity indicates that nearly all moderately probable members are also assigned high-confidence membership, signalling a clean, well-separated cluster/field split for that choice of $k$; lower values indicate a more diffuse, less stable partition. This score, plotted as a function of $k$ in Figure~\ref{fig:upmask_stability} (blue curve), fluctuates from one $k$ to the next because $k$-means partitions are intrinsically sensitive to the initial cluster seeding, but it exhibits a well-defined maximum within the explored range for both clusters. We adopted the $k$ value at which the stability score attains this maximum (marked by the red circles in Figure~\ref{fig:upmask_stability}), namely $k=25$ for NGC 2423 and $k=11$ for NGC 2482, rather than fixing $k$ arbitrarily.

The algorithm was applied to a filtered sample of 90,418 stars for NGC 2423 and 134,090 stars for NGC 2482, considering a limiting magnitude of $G \leq 20.5$ mag. Following a conservative approach, stars with $P \geq 0.5$ were classified as probable members and retained for the subsequent analyses.

\begin{figure}[!h]
    \centering
    \includegraphics[width=0.8\linewidth]{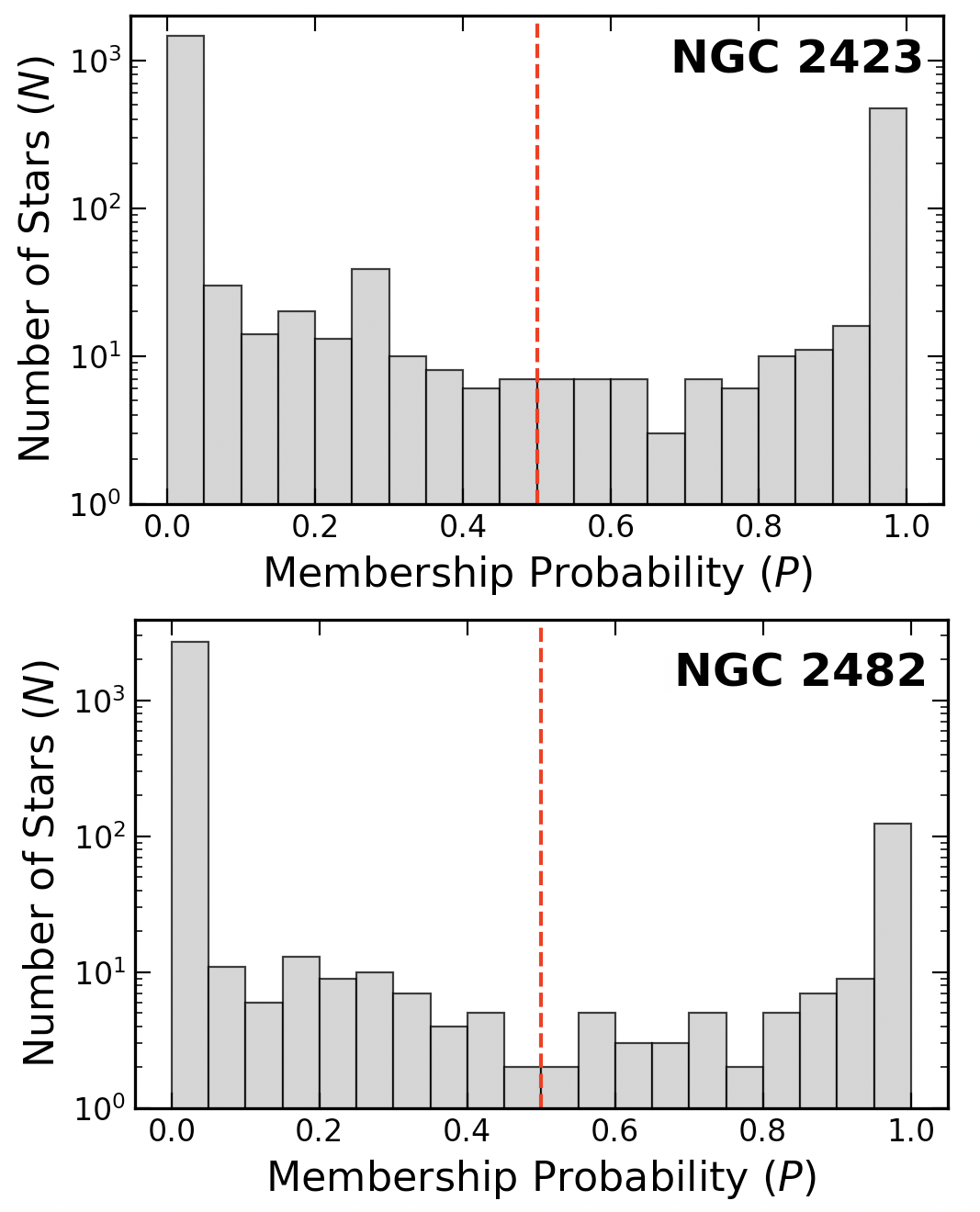}
    \caption{Distribution of membership probabilities ($P$) for NGC 2423 (left) and NGC 2482 (right) obtained with the \textsc{UPMASK} algorithm. The y-axis is shown on a logarithmic scale. The vertical dashed red lines indicate the adopted membership threshold at $P = 0.5$.}
    \label{fig:prob_hist}
\end{figure}

The distribution of membership probabilities ($P$) obtained with the \textsc{UPMASK} algorithm is shown in Figure~\ref{fig:prob_hist}. This diagnostic complements the $k$-selection of Figure~\ref{fig:upmask_stability}: while Figure~\ref{fig:upmask_stability} shows how the sharpness of the field/cluster separation varies with $k$, Figure~\ref{fig:prob_hist} shows that separation directly, for the adopted $k$, in terms of star counts. For both clusters, the distributions are bimodal, with the large majority of (field) stars concentrated at low probabilities ($P < 0.1$), a comparatively low and roughly flat number of stars at intermediate probabilities ($0.1 \lesssim P \lesssim 0.7$), and a second, distinct peak of probable cluster members at high probabilities ($P > 0.9$). The clear separation between these two peaks, with few stars populating the intermediate range, indicates that \textsc{UPMASK} is able to discriminate cluster members from the field population with limited ambiguity for both targets, and it justifies the adopted membership threshold of $P\geq0.5$ (marked by the vertical dashed line) as lying within the low-density trough between the two populations rather than cutting through either peak. The logarithmic $y$-axis is used because the intermediate-probability bins contain far fewer stars than the low- and high-probability peaks; without it these bins would not be visible on a linear scale.

A membership probability threshold of $P \geq 0.5$, indicated by the dashed red line, was adopted to define the initial sample of probable cluster members. This criterion yielded 545 candidate members for NGC 2423 and 164 for NGC 2482. To further improve the reliability of the final samples, an additional quality cut based on the astrometric goodness-of-fit was applied by retaining only stars with \texttt{RUWE} $\leq 1.4$. Following this filtering, the final membership lists consist of 488 stars for NGC 2423 and 116 stars for NGC 2482.

\textcolor{black}{We applied the RUWE~$\leq 1.4$ cut after UPMASK membership assignment, only to the probable members. This approach avoids removing potential cluster members before the membership analysis. UPMASK was first used to identify the kinematically coherent population, followed by the RUWE cut as a quality filter. Our final membership lists contain 488 stars for NGC 2423 and 116 stars for NGC 2482. These values are comparable to those reported in previous \textit{Gaia}-based studies. \citet{Hunt2024} identified 571 and 192 members, while \citet{CantatGaudin2020} reported 406 and 115 members for NGC 2423 and NGC 2482, respectively. The differences are mainly due to the membership probability threshold ($P \geq 0.5$), the RUWE~$\leq 1.4$ quality cut, and differences between the membership methods. Our aim is not to maximize the number of members, but to obtain a homogeneous and quality-controlled sample for the subsequent isochrone fitting and kinematic analysis.}

\textcolor{black}{We note that our choice of UPMASK stems primarily from the need for a homogeneous, cluster-specific membership determination, rather than a categorical claim of superiority over the clustering algorithms underlying large all-sky compilations \citealp[e.g.,][]{Hunt2024, Cantat_2020, Dias_2021}. While such catalogs employ a fixed pipeline configuration across thousands of clusters, the stability score analysis in Figure 1 allowed us to independently adjust the basis free parameter (k-means number of clusters) for each of NGC 2423 and NGC 2482, rather than adopting an arbitrary or externally inherited value. This approach produces similar or larger membership samples compared to most individual literature determinations listed in Table 1 after a RUWE $\leq$ 1.4 quality cutoff, and also offers membership probabilities that can be directly reproduced from the stated methodology; a feature not always achievable from catalog-level membership markers. In this study, the primary rationale for adopting UPMASK is not a direct claim of algorithmic superiority, but rather this combination of cluster-based optimization, quality control, and reproducibility.}

\subsection{Determination of Cluster Center}

\begin{figure*}[!ht]
    \centering
    \includegraphics[width=0.9\linewidth]{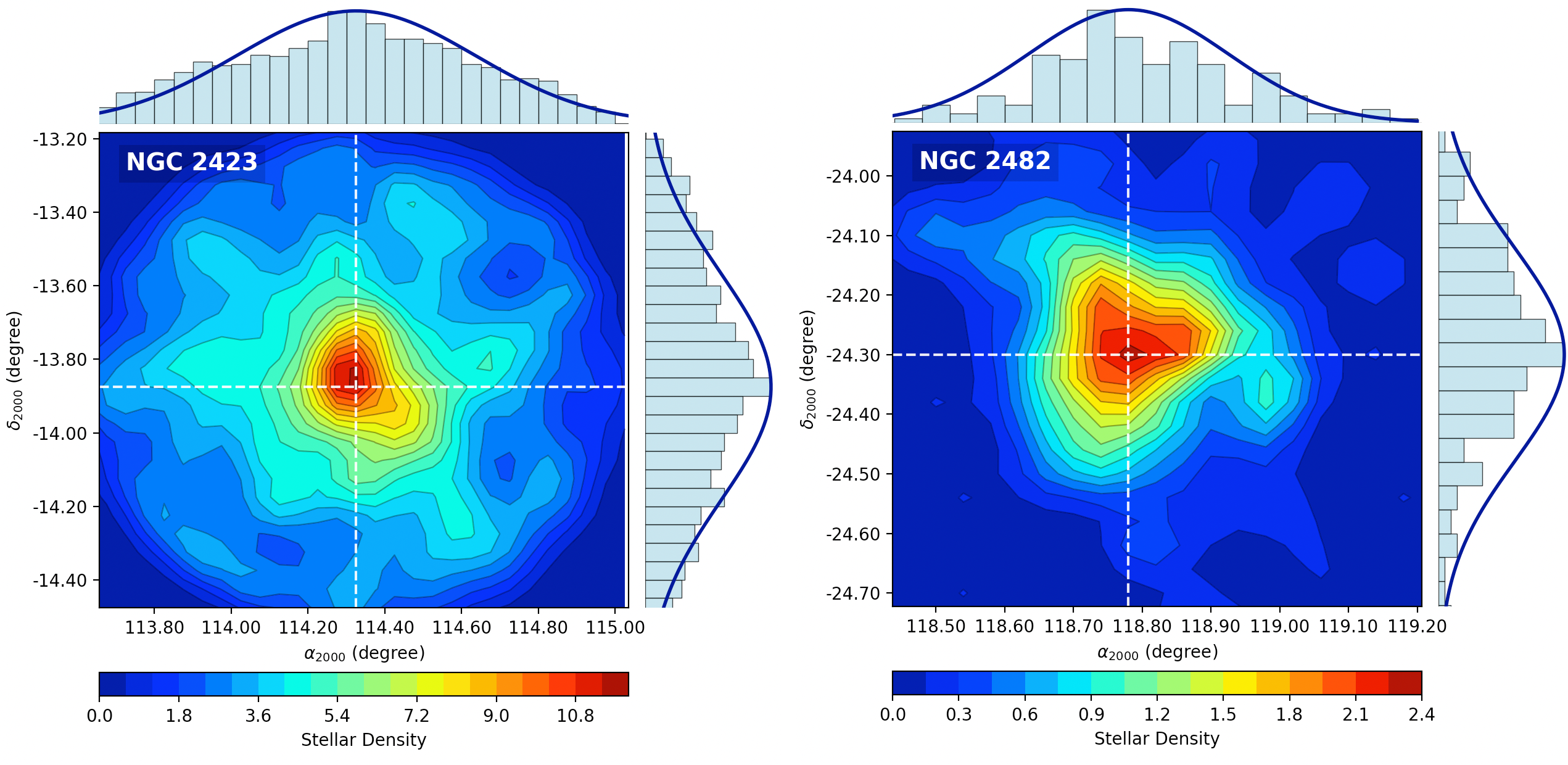}
    \caption{Surface density maps of the open clusters NGC 2423 (left) and NGC 2482 (right) in equatorial coordinates ($\alpha_{\rm J2000}$, $\delta_{\rm J2000}$). The main panels display stellar surface density contour maps, where the colour scale represents the number of stars per unit area. Regions of highest stellar concentration are shown in red, while lower-density areas are indicated in blue. The marginal histograms along the top and right axes show the projected stellar density distributions in $\alpha$ and $\delta$, respectively, together with the corresponding Gaussian fits (blue curves). The dashed lines mark the derived central coordinates of each cluster.}
    \label{fig:centers}
\end{figure*}

The centers of NGC 2423 and NGC 2482 were redetermined using the final lists of probable cluster members. The central coordinates were defined as the regions of maximum stellar surface density in equatorial coordinates. To locate these positions, one-dimensional density distributions in right ascension ($\alpha$) and declination ($\delta$) were constructed from the \textit{Gaia} data. The number of stars within each bin was counted, and Gaussian profiles were fitted to the resulting histograms. The peaks of the Gaussian fits in both $\alpha$ and $\delta$ were adopted as the updated cluster centers. Using this approach, the center of NGC 2423 was determined to be $\alpha = 07^{\rm h}37^{\rm m}17.8^{\rm s}$ and $\delta = -13^\circ52'19.0''$, while the center of NGC 2482 was determined to be $\alpha = 07^{\rm h}55^{\rm m}07.1^{\rm s}$ and $\delta = -24^\circ17'43.3''$. These positions correspond to the highest stellar concentration within each cluster and are illustrated in Figure~\ref{fig:centers}.

\section{Structural Properties of the Clusters}

\begin{figure*}[!ht]
    \centering
    \includegraphics[width=0.5\linewidth]{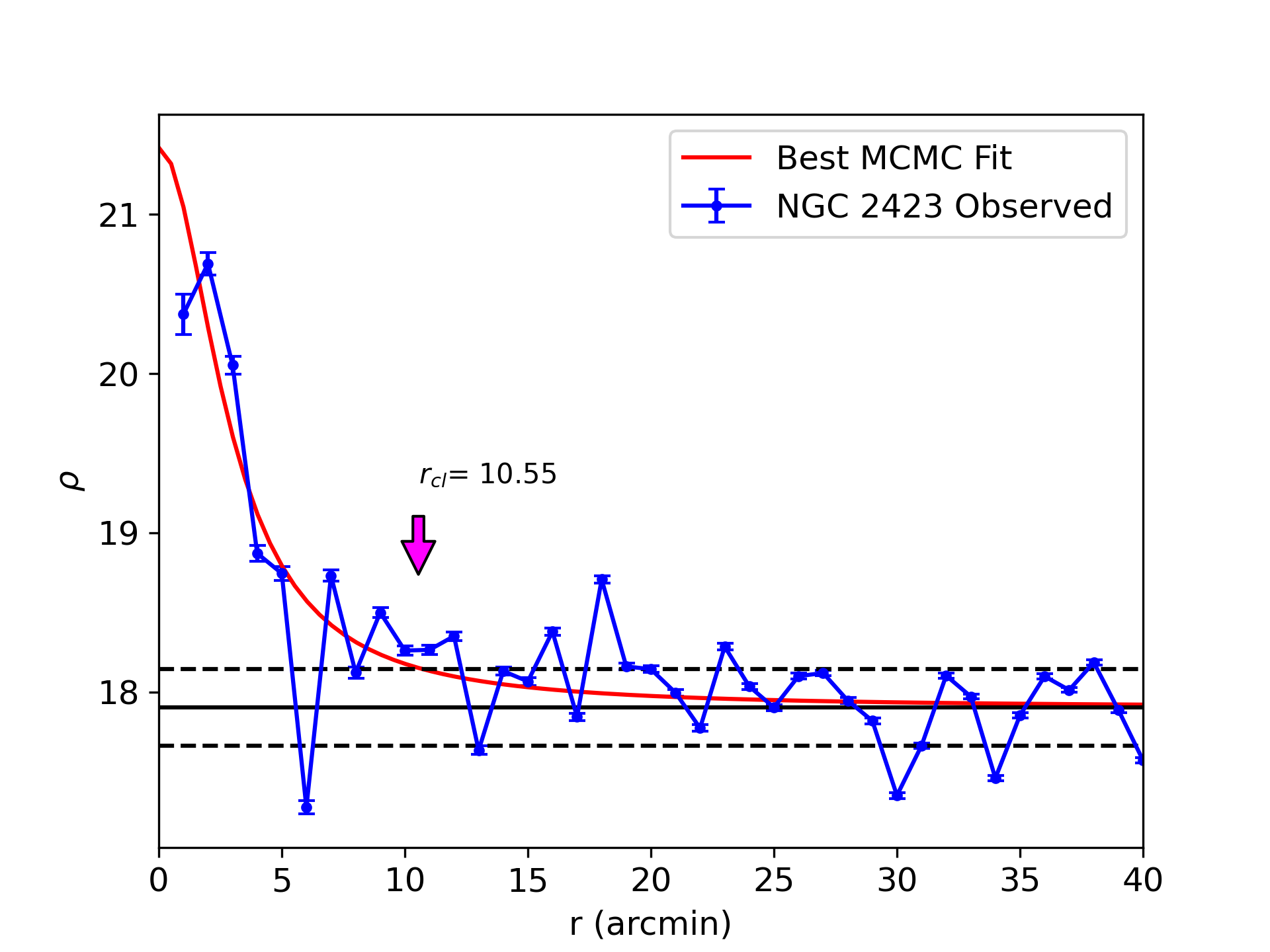}
    \includegraphics[width=0.4\linewidth]{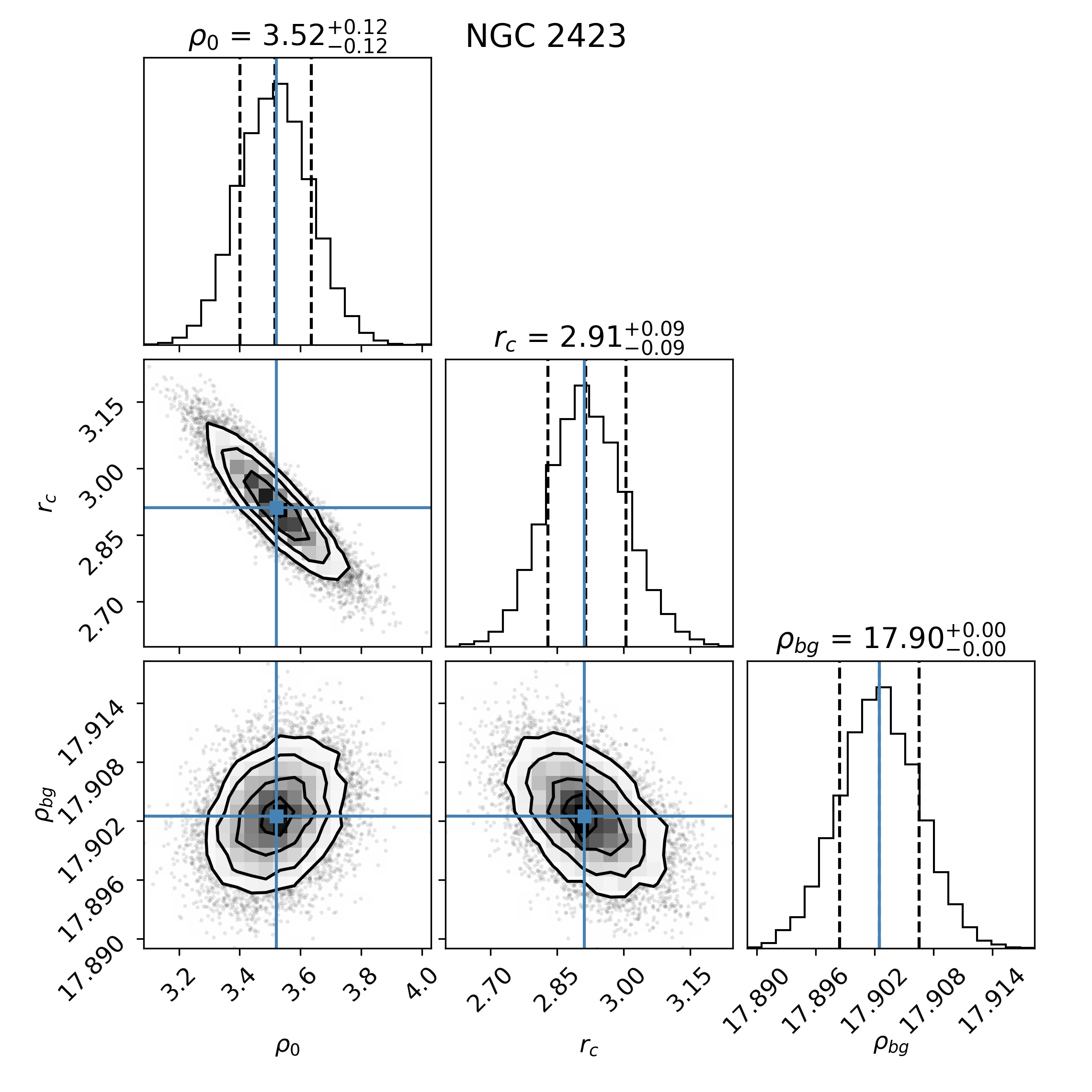}\\
    \includegraphics[width=0.5\linewidth]{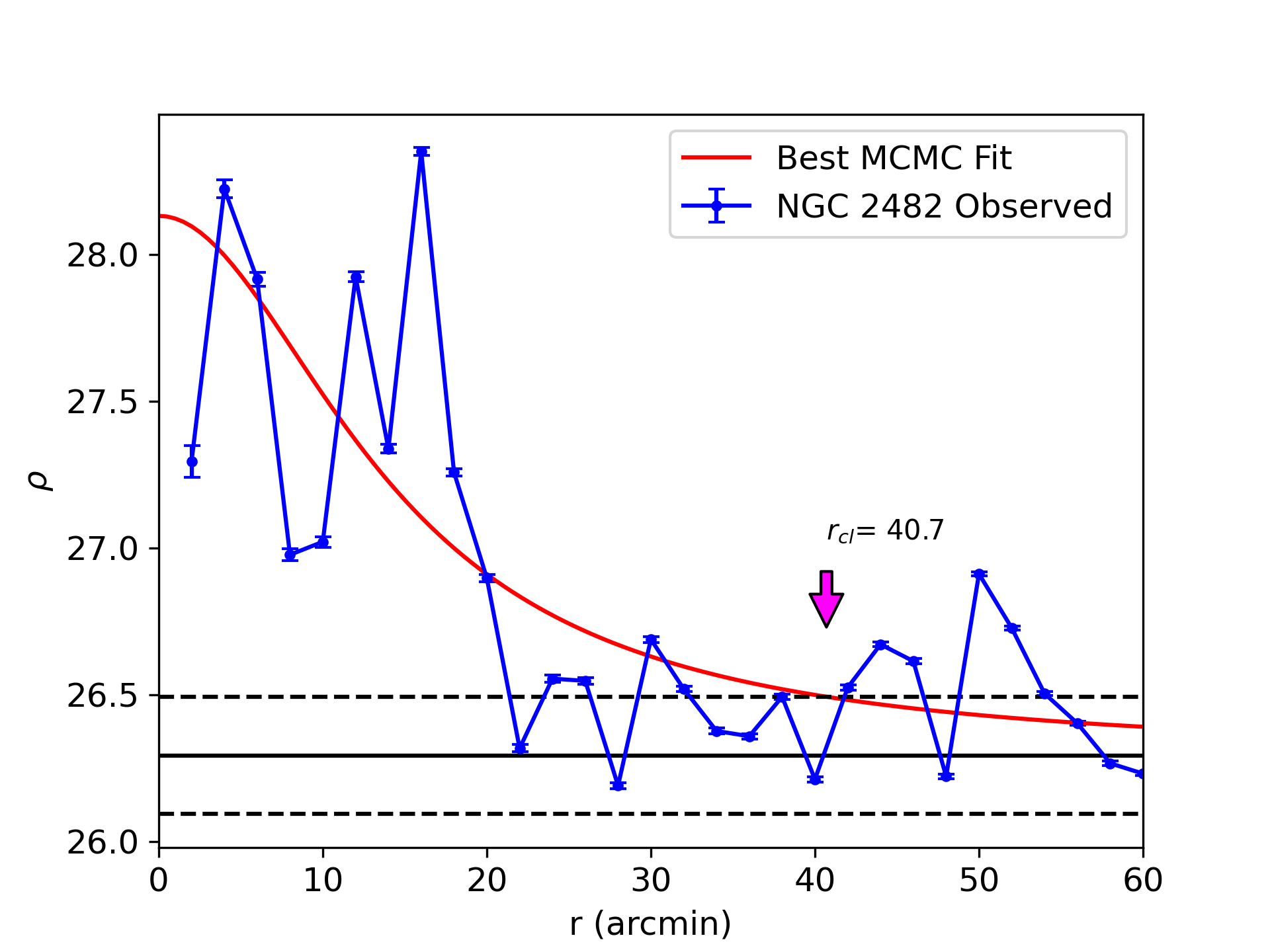}
    \includegraphics[width=0.4\linewidth]{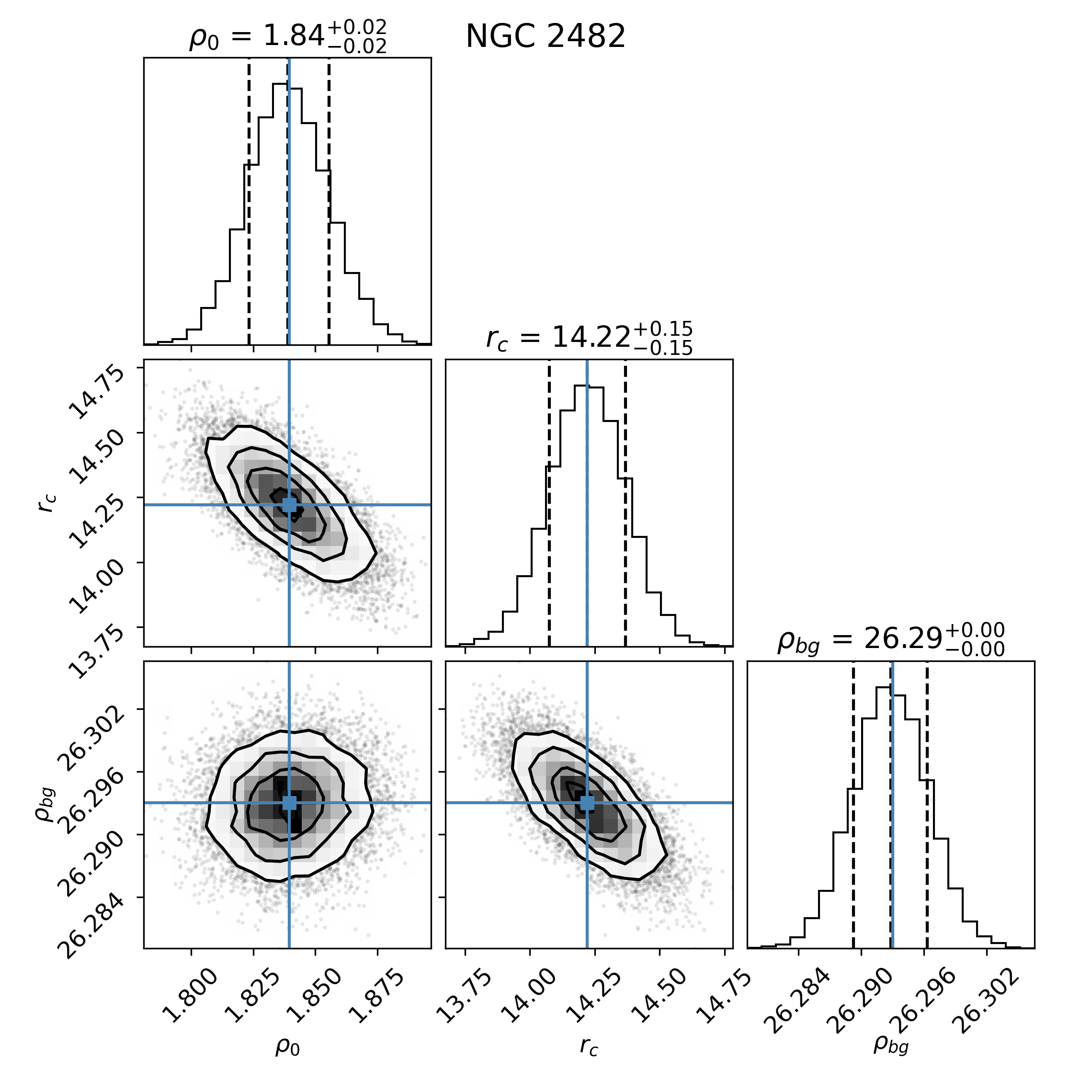}
    \caption{The RDPs of the investigated OCs represented by the blue solid line. The red solid line represents the \citet{King1962} fitting while the black solid line and black dashed line represent the background density $\rho_{\mathrm{bg}}$ and the uncertainties of the background density, respectively.}
    \label{fig:rdps}
\end{figure*}

Radial density profiles (RDPs) provide a simple way to examine the structural properties of open clusters, such as their degree of central concentration and overall richness. For each cluster, the region around the center was divided into concentric annuli, and the stellar surface density in each ring was calculated as
\begin{equation}
\rho(r_i) = \frac{N_i}{A_i},
\label{radial_density}z
\end{equation}
where $N_i$ is the number of stars within the $i$th annulus and $A_i$ is its area. The associated uncertainties were estimated assuming Poisson statistics, $1/\sqrt{N_i}$. The resulting RDPs of NGC 2423 and NGC 2482, together with the best-fitting King models, are shown in Figure~\ref{fig:rdps}. The figure illustrates the observed stellar surface densities as a function of radius, along with the corresponding background levels and model fits for both clusters.

To derive the structural parameters, the empirical surface-density model introduced by \citet{King1962} was fitted to the observed RDPs. We adopted the following form of the model
\begin{equation}
\rho(r) = \rho_{\mathrm{bg}}+\frac{\rho_0}{1+(r/r_c)^2},
\label{eq:king_model_short}
\end{equation}
where $\rho_0$ denotes the central stellar density above the background level, $r_c$ is the core radius, and $\rho_{\mathrm{bg}}$ represents the background density.

The model parameters were estimated using a maximum-likelihood approach, with the log-likelihood defined as
\begin{equation}
\ln \mathcal{L} = -\sum_i
\left( \frac{\rho_i - \rho_{i,\mathrm{model}}}{\sigma_{\rho_i}} \right)^2 .
\end{equation}

Parameter sampling was carried out with the \texttt{emcee} Markov Chain Monte Carlo sampler \citep{Foreman-Mackey2013}, assuming uniform priors. Convergence of the chains was verified using the Gelman–Rubin statistic \citep{gelman1992}. The resulting structural parameters for both clusters are listed in Table~\ref{King_para}.

Following \citet{King1966}, the concentration parameter was computed as $C = r_t / r_c$, which provides a quantitative measure of cluster compactness. \textcolor{black}{The computation of the tidal radii used to get the $C$ parameter is presented later in subsection \ref{sec:internal_dyn}}. The derived concentration values indicate that NGC 2423 is more centrally concentrated than NGC 2482, consistent with the differences seen in their radial density profiles. Our results are generally in agreement with previous studies, with small differences likely arising from variations in the adopted datasets, profile construction methods, and fitting techniques, as well as from the improved quality and homogeneity of the {\it Gaia} DR3 data used in this work.

We note that the observed radial density profiles in Figure~\ref{fig:rdps}, particularly for NGC 2482, do not fully flatten to a stable background level within the extraction radii adopted here ($40'$ for NGC 2423 and $60'$ for NGC 2482), and that the bin-to-bin scatter around the best-fitting King profile is non-negligible, especially at intermediate radii. This indicates that the true tidal radius of NGC 2482 may extend beyond the field of view probed in this work, and that the derived core radius $r_{\rm c}$ and background density $\rho_{\rm bg}$ should be regarded as being subject to this limitation, in addition to the formal MCMC uncertainties quoted in Table~\ref{King_para}. We caution that the structural parameters in Table~\ref{King_para}, and quantities derived from them ($r_{\rm t}$, $R_{\rm h}$, $T_{\rm relax}$; Section~\ref{sec:internal_dyn}), inherit this uncertainty, and that a dedicated extraction over a substantially larger field, allowing the profile to asymptote to the field-star density before the fit, would provide a more robust determination of the tidal structure of NGC 2482 in particular.

\begin{table*}[!h]
\centering
\small
\caption{Derived structural parameters for NGC 2423 and NGC 2482 obtained from fitting the \citet{King1962} model to the radial density profiles. The columns list the cluster radius ($r_{\rm cl}$), core radius ($r_{\rm c}$), central density ($\rho_0$), background density ($\rho_{\rm bg}$), and concentration parameter ($C$).}
\label{King_para}
\renewcommand{\arraystretch}{1.2}
\begin{tabular}{lccccc}
\hline\hline
Cluster & $r_{\rm cl}$ & $r_{\rm c}$ & $\rho_0$ & $\rho_{\rm bg}$ & $C$  \\
\cline{2-3} \cline{4-5}
 & \multicolumn{2}{c}{(arcmin $\&$ pc)} & \multicolumn{2}{c}{(stars arcmin$^{-2}$)} &    \\
\hline
NGC 2423 & 10.55 $\&$ 2.90  & 2.91$^{+0.09}_{-0.08}$ $\&$ 0.80 $\pm$ 0.04 &  $3.52^{+0.11}_{-0.12}$  & $17.90^{+0.004}_{-0.004}$   &  \textcolor{black}{13.91} \\
NGC 2482 & 40.70 $\&$ 16.12 & 14.23$^{+0.15}_{-0.14}$ $\&$ 5.64 $\pm$ 2.37 & $1.84^{+0.02}_{-0.02}$  & $26.29^{+0.003}_{-0.003}$ &  \textcolor{black}{1.35} \\
\hline
\end{tabular}
\end{table*}

\textcolor{black}{We also examined the radial distribution of the 188 NGC 2423 members with available radial-velocity measurements. The distribution is broadly consistent with that of the full UPMASK member sample, indicating that the spectroscopic members trace the central concentration of the cluster. However, the radial-velocity sample is not spatially homogeneous, particularly at large cluster-centric radii, because spectroscopic observations are generally biased toward brighter and more easily observable stars. Consequently, structural parameters derived solely from this subsample are not sufficiently robust to replace those obtained from the full membership sample.}

\section{Astrometric Properties of the Clusters}

\begin{figure}[h!]
\centering
\includegraphics[width=0.99\linewidth]{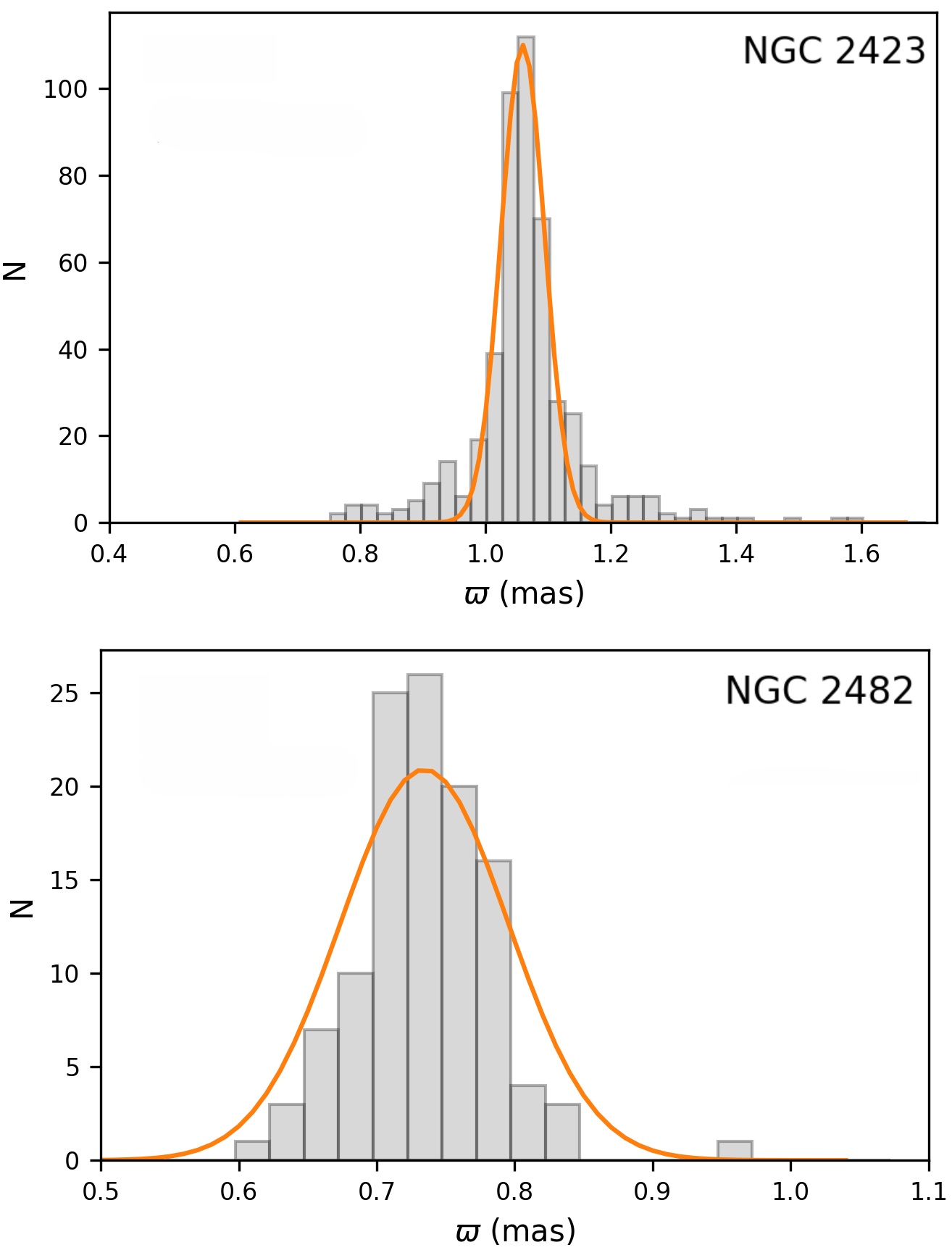}
\caption{Histograms of the trigonometric parallaxes for all member stars of the clusters investigated in this work. The mean values and standard deviations are shown in each panel, while the red curves represent the Gaussian fits to the observed trigonometric parallax distributions.
\label{fig: Plx}}
\end{figure}

\begin{figure*}[htb!]
\captionsetup[subfigure]{labelformat=empty}
\centering
\subfloat[]{\scalebox{.45}{\includegraphics{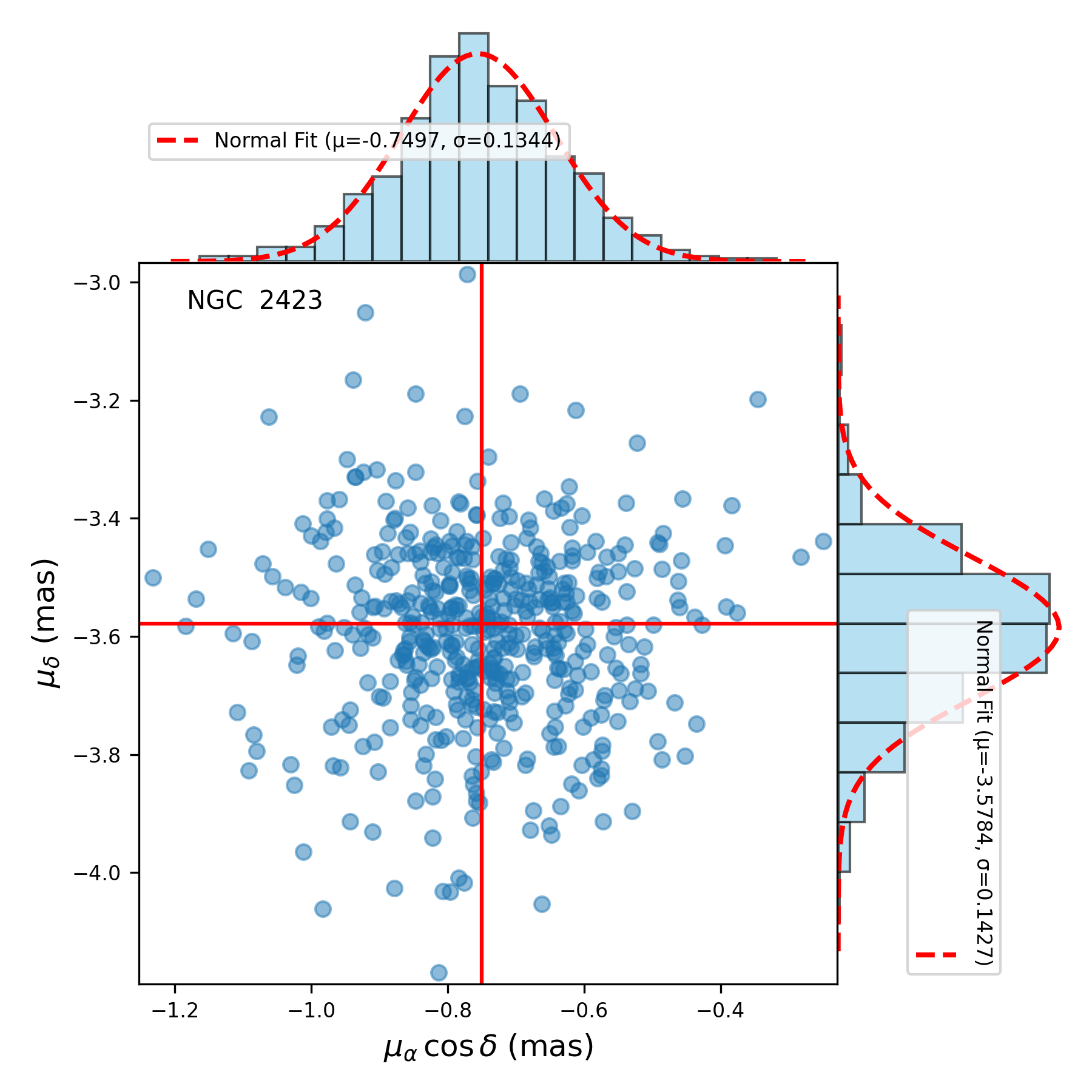}}}
\subfloat[]{\scalebox{.45}{\includegraphics{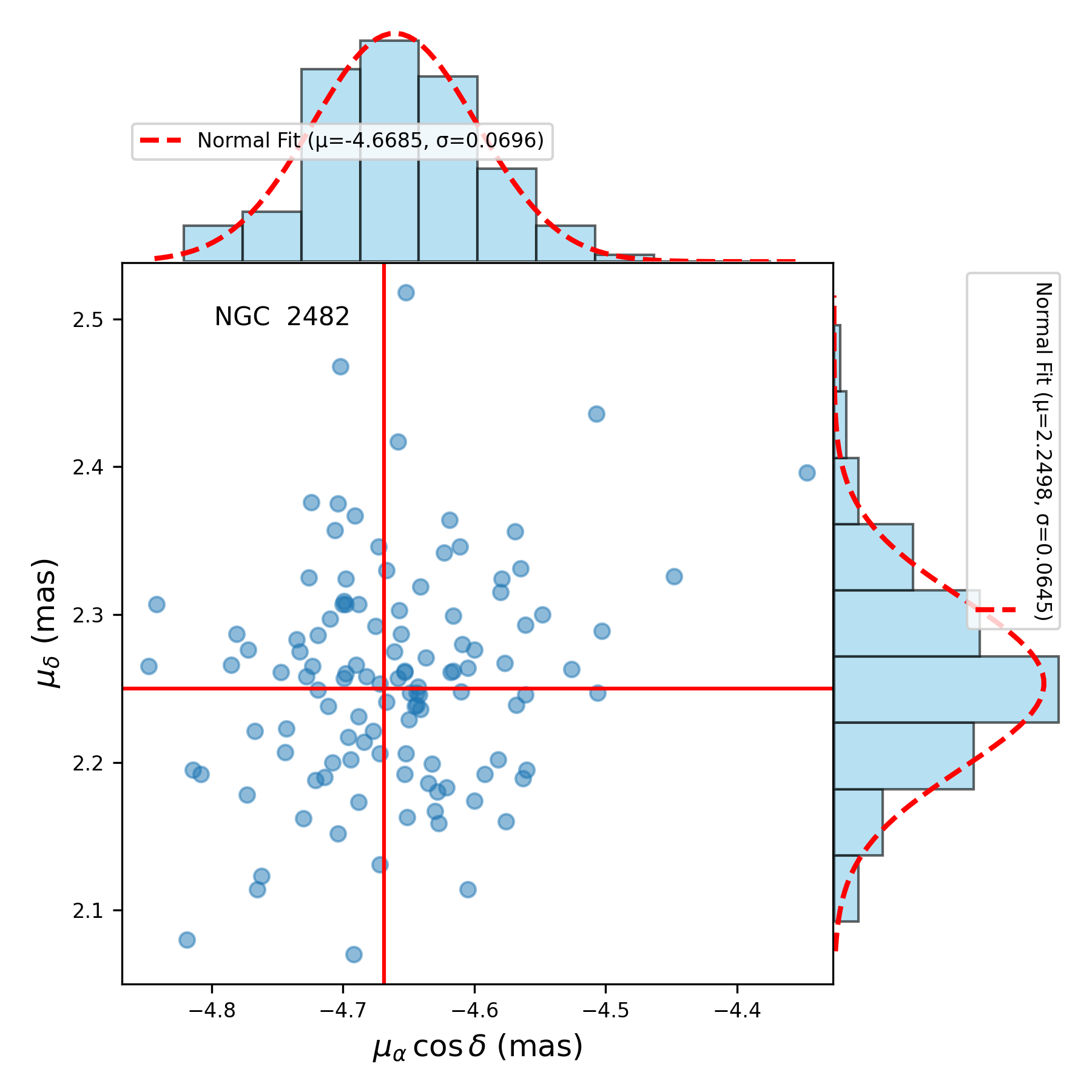}}}
\caption{Histograms of the proper motion components ($\mu_{\alpha}\cos\delta$,~$\mu_{\delta}$) for the member stars of each target cluster. The mean value and standard deviation for each distribution are noted in the panels.
\label{fig: PM}}
\end{figure*}

The mean proper-motion components in right ascension and declination
($\mu_{\alpha}\cos\delta$ and $\mu_{\delta}$), as well as the mean trigonometric parallaxes ($\varpi$), were estimated for both clusters using the astrometric data of the selected member stars. For each cluster, histograms of the proper-motion components and parallaxes were constructed, and Gaussian functions were fitted to the observed distributions. The mean values and standard deviations of the best-fitting Gaussians were adopted as the representative astrometric parameters of the clusters (see Figures~\ref{fig: Plx} and \ref{fig: PM}).

The mean proper-motion and trigonometric-parallax values for NGC 2423 and NGC 2482 were derived from the member stars. The distributions and Gaussian fits are illustrated in Figures~\ref{fig: PM} and \ref{fig: Plx}, and the resulting numerical values are listed in Table~\ref{all_results}. As a consistency check, we compared our astrometric results with the compilation in Table~\ref{tab:literature}. As expected for a homogeneous, single-catalogue analysis, our values agree closely with other studies based on the same {\it Gaia} astrometric solutions \citep[e.g.][]{Hunt2024, Poggio_2021, Dias_2021}, since these share the same underlying parallax and proper-motion zero points and comparable precision. The larger discrepancies with pre-{\it Gaia} compilations \citep[e.g.][]{sampedro2017, Dias2014, kharchenko2013vizier} (see Table~\ref{tab:literature}) are expected rather than indicative of any inconsistency in our analysis: those catalogues relied on ground-based proper motions (e.g., UCAC4) with substantially larger and less well-characterized astrometric uncertainties than {\it Gaia} DR3. We report this comparison only to confirm that our membership and mean-parameter determination reproduce the expected {\it Gaia}-era values, not as an independent astrometric result.

Astrometric distances were obtained from the mean cluster parallax through the simple inverse relation $d_{\varpi}=1/\overline{\varpi}$ (with $\overline{\varpi}$ in arcsec, $d_{\varpi}$ in pc), using the mean and standard deviation of the Gaussian fitted to the parallax distribution of the final member sample (Figure~\ref{fig: Plx}); the associated uncertainty was propagated from $\sigma_{\overline{\varpi}}$ via $\sigma_{d_\varpi} = d_\varpi^2\,\sigma_{\overline{\varpi}}$. This yields $944 \pm 31$ pc for NGC 2423 and $1361 \pm 113$ pc for NGC 2482. Given the modest relative parallax uncertainties of the adopted samples ($\sigma_{\overline{\varpi}}/\overline{\varpi}\lesssim7\%$), this direct inversion is not expected to introduce a significant bias relative to a full Bayesian distance estimate, and the resulting distances are consistent, within uncertainties, with values reported in the literature based on the most recent {\it Gaia} releases \citep[e.g.][]{Poggio_2021, Dias_2021}. 

\textcolor{black}{We note that the parallax distribution of NGC 2423 shows a small excess of stars beyond $\varpi \sim 1.4$~mas. We examine the spatial and proper-motion distributions of the member stars of both clusters as a function of parallax in Appendix~\ref{app:parallax_tail} and show that these high-parallax NGC 2423 sources are dispersed relative to the bulk membership sample, both spatially and kinematically, consistent with residual field-star contamination rather than a distinct population of closer members; their contribution to the derived mean parallax and distance is negligible ($<0.005$~mas, well within the quoted uncertainty). No comparable tail or outlier population is present for NGC 2482.}

\begin{figure*}[!h]
    \centering
\includegraphics[width=0.85\linewidth]{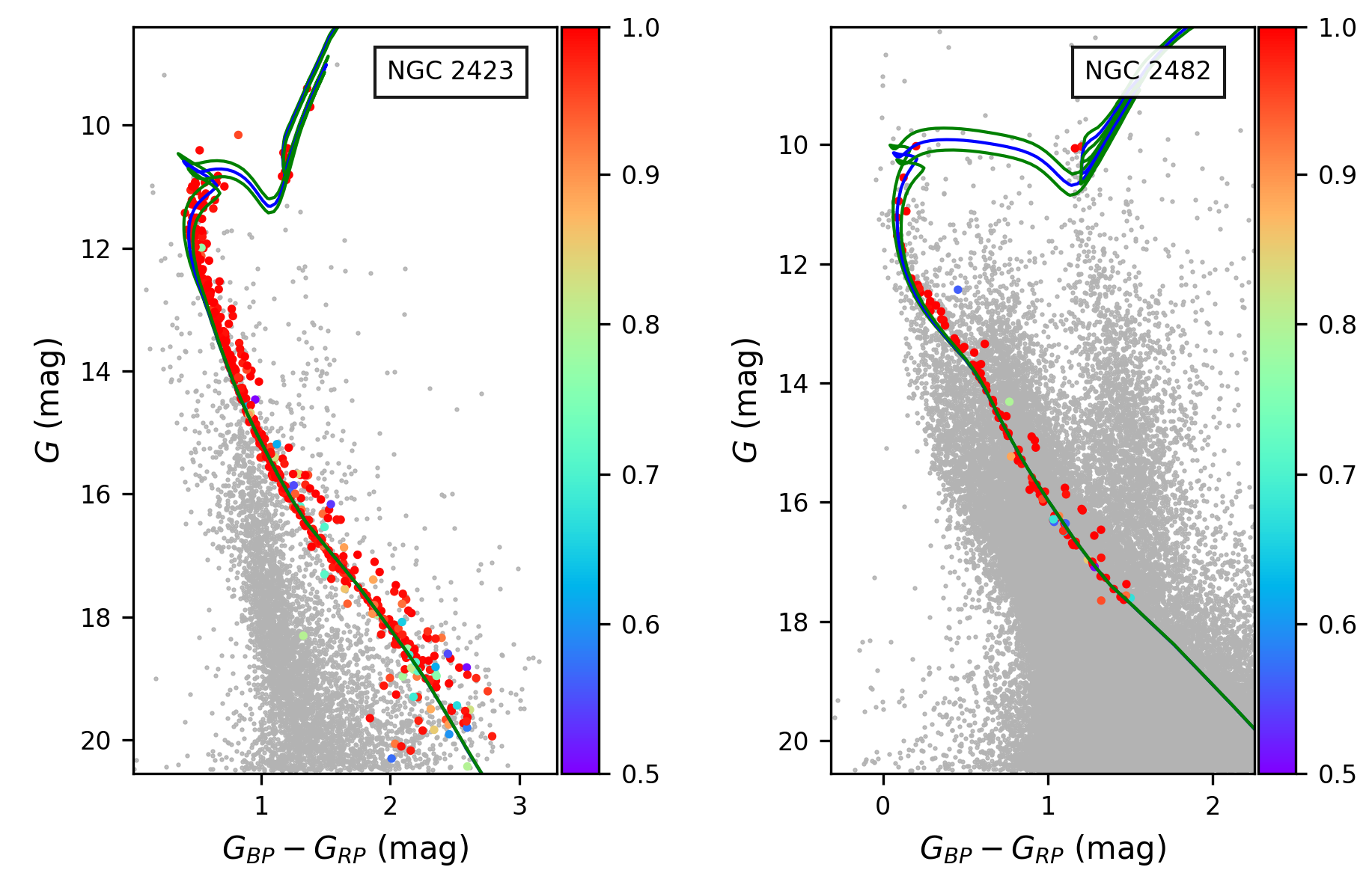}\\
\includegraphics[width=0.85\linewidth]{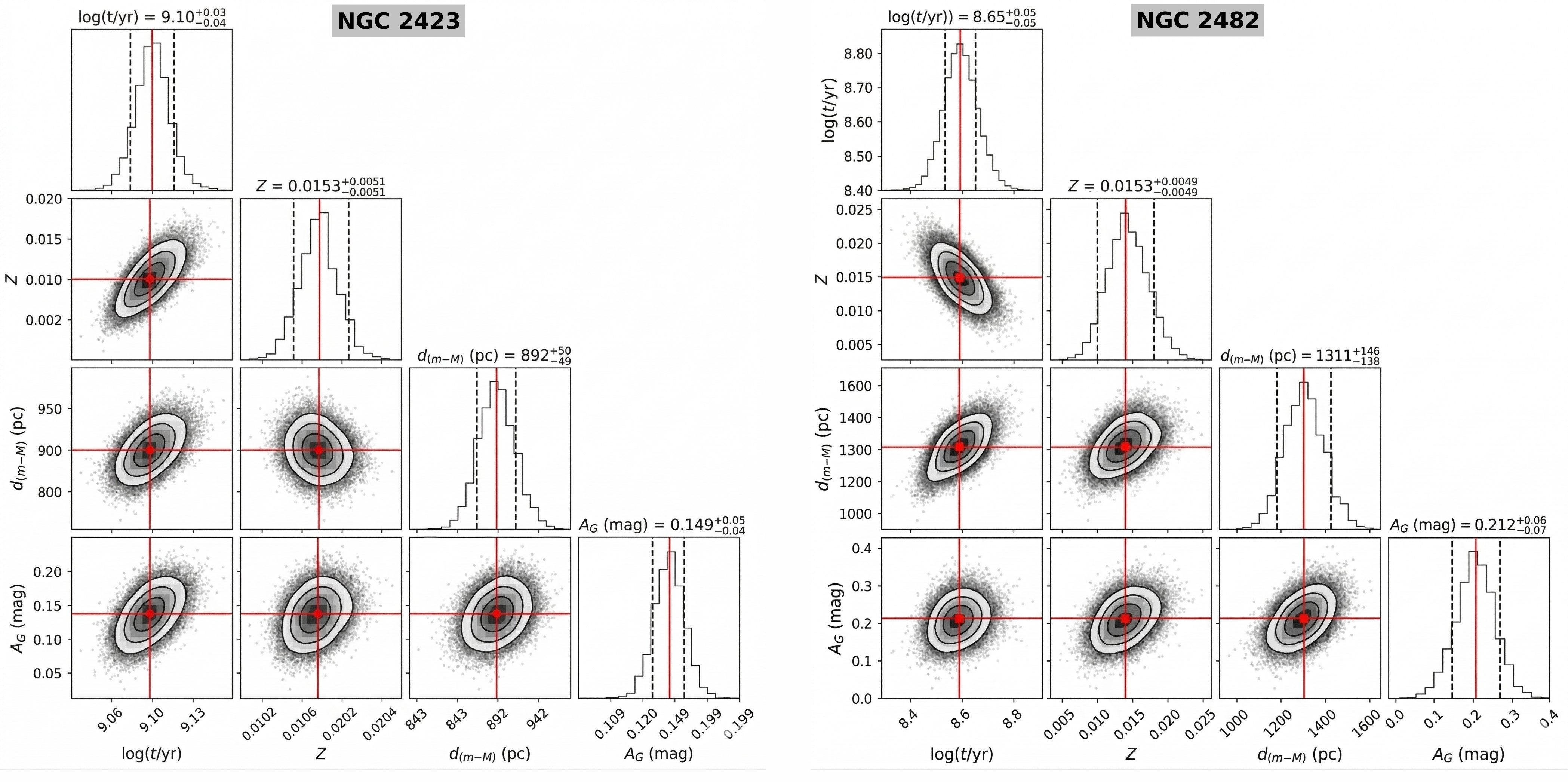}
    \caption{CMDs of the open clusters NGC 2423  and NGC 2482. Probable cluster members with $0.5 \leq P \leq 1$ are shown in distinct colour codes, while field stars are represented by grey circles. The best-fitting {\sc PARSEC} isochrones are represented by blue lines, with their corresponding uncertainty ranges represented by green lines.}
    \label{CMDs}
\end{figure*}

\textcolor{black}{As an independent check on the distances obtained from direct
trigonometric parallax inversion, we cross-matched our final member lists with the Bayesian distance catalogue of \citet{BailerJones2021}, which provides geometric and photogeometric distance estimates ($r_{\rm geo}$, $r_{\rm pgeo}$) for {\it Gaia} DR3 sources based on a probabilistic inference incorporating a direction-dependent prior on the stellar density distribution of the Galaxy. Taking the mean photogeometric distance of the member stars, we obtain $r_{\rm pgeo} = 931 \pm 68$~pc for NGC 2423 and $r_{\rm pgeo} = 1311 \pm 56$~pc for NGC 2482. These values are in excellent agreement with our parallax-inversion distances ($944 \pm 31$~pc and $1361 \pm 113$~pc, respectively) and with our isochrone-fitting photometric distances ($892 \pm 50$~pc and $1311 \pm 146$~pc, respectively), confirming that the direct inversion $d_\varpi = 1/\varpi$ does not introduce a significant bias for either cluster, consistent with the modest relative parallax uncertainties of the adopted samples ($\sigma_\varpi/\varpi \lesssim 7\%$).}

\section{Astrophysical Properties of the Clusters}

The fundamental astrophysical parameters of NGC 2423 and NGC 2482 were derived from their observed CMDs, constructed using {\it Gaia} photometry for the selected cluster members. The observed CMDs were fitted with theoretical stellar evolutionary models to determine the best-matching cluster properties.

In this study, we adopted {\sc PARSEC} isochrones \citep{Bressan2012}, computed for a Solar metallicity of $Z_{\odot} = 0.0152$. For clusters with available iron abundance measurements, the reported [Fe/H] values were converted into the corresponding heavy-element mass fraction $Z$ to ensure consistency with the metallicity scale of the {\sc PARSEC} models. This conversion follows the analytical prescriptions calibrated for the PARSEC framework \footnote{\url{https://github.com/jobovy/isodist/blob/master/isodist/Isochrone.py}}. The cluster parameters were fitted using a Markov Chain Monte Carlo (MCMC) approach, following the methodology of \citet{Bilir2026}. The exploration of the posterior space was performed with the affine-invariant ensemble sampler in the \texttt{emcee} package \citep{Foreman-Mackey2013}, with the number of walkers scaled to the number of member stars and each chain run for 5000 steps. After discarding the burn-in phase, the posterior distributions were analyzed to obtain median values and the $1\sigma$ confidence intervals (16th–84th percentiles).The heavy-element mass fraction $Z$ was calculated from the measured [Fe/H] values following the standard prescriptions \citep{Yontan2023,2023AJ....165...79Y, Haroon2025, Elsanhoury2025}:
\begin{equation}
z_{\rm x} = 10^{{\rm [Fe/H]} + \log\left(\frac{Z_{\odot}}{1 - 0.248 - 2.78~Z_{\odot}}\right)},
\end{equation}
\begin{equation}
Z = \frac{z_{\rm x} (1 - 0.2485)}{1 + 2.78~z_{\rm x}}.
\end{equation}
These relations provide a consistent conversion between iron abundance and heavy-element mass fraction, ensuring uniformity in the isochrone fitting procedure across the sample \citep{ Cinar2026c, Cinar2026d, TasdemirCinar2025, Tasdemir2026raa, Canbay2026b}. Neither NGC 2423 nor NGC 2482 has a spectroscopic [Fe/H] determination based on high-resolution data for a statistically robust sample of members; the only literature value listed in Table~\ref{tab:literature}, ${\rm [Fe/H]}=0.03\pm0.14$~dex for both clusters \citep{Dias_2021}, is a photometric proxy calibrated against a reference catalogue of clusters with spectroscopic abundances, rather than a direct spectroscopic measurement, and carries an uncertainty comparable to its own value. Given the absence of a reliable, cluster-specific spectroscopic abundance, and since this proxy is statistically indistinguishable from Solar, we adopted a Solar metallicity of [Fe/H] = 0.00 dex ($Z=0.0152$) for the isochrone fitting of both clusters, consistent with the typical metallicity of the young-to-intermediate-age thin-disc population at the Galactocentric radii of NGC 2423 and NGC 2482 ($R_{\rm GC}\approx8.8$--$8.9$~kpc; Table~\ref{all_results}). We caution that this assumption, rather than a measured value, is a potential source of systematic uncertainty in the derived ages and distances, in addition to the quoted statistical errors.

Figure~\ref{CMDs} presents the $G$ versus $(G_{\rm BP}-G_{\rm RP})$ CMDs of the analysed clusters, together with their best-fitting isochrones. The isochrone data provide absolute magnitudes ($M_G$) and intrinsic colours as a function of stellar mass at a fixed age. These quantities were used to estimate the observed distance modulus and colour excess of each cluster. The observed distance modulus was obtained as $m_G - M_G$, while the colour excess was derived as $E(G_{\rm BP}-G_{\rm RP})$, defined as the difference between the observed stellar colours and the intrinsic isochrone colours.

The derived astrophysical parameters from the isochrone fitting, including ages, distance moduli, reddening, and distances, are summarized in Table~\ref{all_results}.  The colour excesses were computed from the {\it Gaia} colours using $E(B-V) = 0.775 \times E(G_{\rm BP}-G_{\rm RP})$ \citep{cardelli1989relationship}, and the corresponding extinction in the $G$ band was estimated as $A_G = 2.74 \times E(B-V)$ \citep{casagrande2018use, zhong2019substructure, 2023AJ....165..163C}. The uncertainties on $E(B-V)$ and $A_G$ listed in Table~\ref{all_results} were propagated directly from the fitted $E(G_{\rm BP}-G_{\rm RP})$ uncertainty through these linear coefficients. Photometric distances were derived from the distance modulus using standard relations. \textcolor{black}{Since the derived ages and distances from the observed CMDs are dependent on the metallicity of adopted isochrones, and many previous works suggest a higher metallicity of the OC NGC 2423 \citep{2024ApJ...968...22B,Dias_2021}, we refitted the observed CMD of the OC NGC 2423 with isochrones computed for a higher metallicity of 0.0215 or [Fe/H] $= +0.15$. We found that the best fit isochrone for the higher metallicity has an age of 1295 Myr and the derived distance is 933 pc which are not significantly different from the derived values assuming a solar metallicity.} 

A comparison with values reported in the literature shows that our age estimates are consistent with those of \cite{Poggio_2021}, \cite{Dias_2021}, \cite{CantatGaudin2020}, and \cite{kharchenko2013vizier}, while they are younger than those reported by \cite{Hunt2024}. The age listed for \citet{sampedro2017} in Table~\ref{tab:literature} is lower than our value; we note that \citet{sampedro2017} is primarily a UCAC4-based membership catalogue and does not perform an independent isochrone-fitting age determination, so the age quoted for it in Table~\ref{tab:literature} is a cross-matched literature value (originally traceable to the \citealt{kharchenko2013vizier} MWSC compilation) rather than a result derived from that study's own data. We retain it in Table~\ref{tab:literature} for completeness but do not treat it as an independent age constraint in this comparison.

The Galactic positions of the clusters, including their heliocentric Cartesian coordinates ($X_{\odot}$, $Y_{\odot}$, $Z_{\odot}$) and Galactocentric distances ($R_{\rm GC}$), were calculated following the methodology of \citet{Bisht2026b, Bisht2026c}. We adopted $R_{\odot} = 8.20 \pm 0.10$ kpc \citep{bland2019galah} for consistency with our companion Galactic-dynamics studies \citep[e.g.,][]{Cinar2025, Cinar2026b, Alzahrani2025a, Alzahrani2025b}, which use the same value. This choice lies within $\sim1\sigma$ of other commonly adopted recent determinations of $R_\odot$ (which cluster around 8.1--8.2~kpc), and the resulting sub-percent-level differences in $R_\odot$ are negligible compared to the $\gtrsim$1--3\% uncertainties already present in the cluster distances and space velocities used here. All resulting values are summarized in Table~\ref{all_results}.

\begin{table}[ht]
\centering
\caption{Summary of the results obtained in the present work.}
\label{all_results}
\scalebox{0.85}{
\begin{tabular}{lcc}
\hline
Parameter & NGC 2423 & NGC 2482 \\
\hline
\hline
$\alpha$ (hh:mm:ss)  & $\rm 7^h37^m17.80^s$  & $\rm 7^h55^m07.10^s$ \\
$\delta$ (dd:mm:ss) & $\rm -13^{\circ}52'19.00''$ & $\rm -24^{\circ}17'43.30''$ \\
$(l,~b)^{o}$ & 230.483, 3.585 & 241.648, 2.016 \\
\hline
\multicolumn{3}{c}{Astrometric Parameters} \\
\hline
$\varpi$ (mas) & 1.060 $\pm$ 0.035 & 0.735 $\pm$ 0.061 \\
$d_{\varpi}$ (pc) & 944 $\pm$ 31 & 1361 $\pm$ 113\\
\textcolor{black}{$d_{\rm pgeo}$ (pc)} & \textcolor{black}{931 $\pm$ 68} & \textcolor{black}{1311 $\pm$ 56} \\
$\mu_{\alpha}\cos\delta$ (mas\,yr$^{-1}$) & -0.750 $\pm$ 0.134 & -4.669 $\pm$ 0.070 \\
$\mu_{\delta}$ (mas\,yr$^{-1}$) & $-3.578\pm0.143$ & $2.250\pm0.065$ \\
$N_{\rm stars}$ & 488 & 116 \\
\hline
\multicolumn{3}{c}{Astrophysical Parameters} \\
\hline
$(m-M)_{\rm obs}$ (mag) & $9.90\pm0.10$ & $10.80\pm0.20$ \\
$E(G_{\rm BP}-G_{\rm RP})$ (mag) & $0.07\pm0.01$ & $0.10\pm0.02$ \\
$E(B-V)$ (mag) & $0.054\pm0.008$ & $0.078\pm0.016$ \\
$A_{G}$ (mag) & $0.149\pm0.021$ & $0.212\pm0.042$ \\
$d_{(m-M)}$ (pc) & $892\pm50$ & $1311\pm146$ \\
$Z$ (fixed, Solar) & 0.0152 & 0.0152 \\
$[{\rm Fe/H}]$ (dex, fixed, Solar) & 0.00 & 0.00 \\
Age (Myr) & $1250\pm100$ & $445\pm50$ \\
$\log~({\rm t/yr})$ & $9.10^{+0.03}_{-0.04}$ & $8.65^{+0.05}_{-0.05}$ \\
\hline
\multicolumn{3}{c}{Distances from the Galactic Plane and Centre} \\
\hline
$X_{\odot}$ (kpc) & $-0.599$ & $-0.649$ \\
$Y_{\odot}$ (kpc) & $-0.727$ & $-1.202$ \\
$Z_{\odot}$ (kpc) & 0.059 & 0.048 \\
$R_{\rm GC}$ (kpc) & 8.830 & 8.930 \\
\hline
\end{tabular}}
\end{table}

\section{Kinematic and Dynamical Analysis}

\subsection{Determination of Mean Cluster Radial Velocities}\label{sec:mean_rv}

Radial velocity measurements play a key role in probing the kinematic properties and dynamical histories of open clusters. In this work, the mean radial velocities $\langle V_{\rm r} \rangle$ of NGC 2423 and NGC 2482 were derived using {\it Gaia} DR3 data \citep{GaiaDR3}. The cluster membership selection had already incorporated quality cuts on the astrometric solutions, including a filtering based on \texttt{RUWE} values, ensuring the reliability of the adopted stellar sample.

Only stars classified as cluster members were considered in the radial velocity analysis. Following the methodology described by \citet{Soubiran_2018}, the mean radial velocity of each cluster was computed as an inverse-variance-weighted mean of the individual stellar $V_{\rm rad}$ values, with weights $w_i = 1/\sigma_{V_{{\rm rad},i}}^2$, so that stars with smaller individual radial-velocity uncertainties contribute proportionally more to $\langle V_{\rm r}\rangle$; the reported uncertainty on $\langle V_{\rm r}\rangle$ is the corresponding standard error of the weighted mean, $\sigma_{\langle V_{\rm r}\rangle} = \left(\sum_i w_i\right)^{-1/2}$. For NGC 2423, radial velocity measurements were available for 188 member stars, yielding $\langle V_{\rm r} \rangle = 18.54$ km~s$^{-1}$ with $\sigma_{\langle V_{\rm r}\rangle} = 0.08$ km~s$^{-1}$. In the case of NGC 2482, 29 member stars were used, resulting in $\langle V_{\rm r} \rangle = 39.04$ km~s$^{-1}$ and $\sigma_{\langle V_{\rm r}\rangle} = 0.14$ km~s$^{-1}$. The unweighted (arithmetic) mean and standard deviation of the same samples agree with these weighted values to within their respective uncertainties, confirming that the result is not driven by a small number of high-weight stars.

The {\it Gaia} source identifiers, equatorial coordinates, $G$-band apparent magnitudes, radial velocities, \texttt{RUWE} values, and membership probabilities of the stars used in these calculations are presented in Table~\ref{tab:RV_info}. Only a representative subset of the data is shown in the table, while the complete dataset for both clusters is provided as supplementary online material.

The derived $\langle V_{\rm r} \rangle$ are fully consistent with previous measurements reported in the literature for both clusters. For NGC 2423, published radial velocity estimates span the range $\sim$18.5–21.7 km~s$^{-1}$, with the majority of high-quality determinations clustering around $\sim$19 km~s$^{-1}$ (e.g., \citealt{Tarricq_2021, Dias_2021, Soubiran_2018}). Our result, $\langle V_{\rm r} \rangle = 18.54\pm0.08$ km~s$^{-1}$, lies well within this distribution and is in particularly close agreement with recent {\it Gaia}-based studies, while benefiting from a significantly larger number of member stars and a correspondingly smaller uncertainty.

A similar level of agreement is found for NGC 2482. Previous studies report mean radial velocities mostly concentrated around $\sim$39 km~s$^{-1}$, with occasional outliers associated with larger uncertainties or lower-quality measurements (e.g., \citealt{Carraro_2017}). The value obtained in this work, $\langle V_{\rm r} \rangle = 39.04\pm0.14$ km~s$^{-1}$, is fully consistent with recent high-precision determinations based on {\it Gaia} DR3 data \citep{Tarricq_2021, Dias_2021}. The close agreement across independent studies confirms the robustness of our membership selection and weighted mean approach, and supports the reliability of the adopted radial velocities for subsequent kinematic and orbital analyses.

\begin{table*}
\centering
\footnotesize
\renewcommand{\arraystretch}{0.9}
\setlength{\tabcolsep}{3pt}
\caption{Representative entries from the complete membership catalogues of NGC~2423 and NGC~2482. The complete membership catalogues are provided as supplementary online material.}
\label{tab:membership_catalogues}
\begin{tabular}{r l c c c c c c c c c c}
\toprule
No. & $Gaia$ DR3 ID & $\alpha$ & $\delta$ (deg) & $G$ (mag) & $(G_{\rm BP}-G_{\rm RP})$ & $\mu_{\alpha}\cos\delta$ & $\mu_{\delta}$ & $\varpi$ & $V_{\rm rad}$ & RUWE & $P_{\rm mem}$ \\
 & &(deg) &(deg) & (mag) & (mag) & (mas yr$^{-1}$) & (mas yr$^{-1}$) & (mas) & (km s$^{-1}$) & & \\
\midrule
\multicolumn{12}{c}{NGC 2423} \\
\hline
1 & 3030268653345162368 & 114.2838 & -13.8589 & 15.91 & 1.19 & -0.75 & -3.57 & 1.08 & --- & 1.04 & 1.00 \\
2 & 3030269203100923264 & 114.3377 & -13.8241 & 17.94 & 2.16 & -0.83 & -3.82 & 0.97 & --- & 1.05 & 1.00 \\
3 & 3030267893128871680 & 114.3889 & -13.8691 & 14.06 & 0.78 & -0.46 & -3.55 & 1.04 & 16.71 & 0.99 & 0.99 \\
\ldots & \ldots & \ldots & \ldots & \ldots & \ldots & \ldots & \ldots & \ldots & \ldots & \ldots & \ldots \\
488 & 3030260544447025920 & 114.1812 & -13.9624 & 11.35 & 0.55 & -0.67 & -3.59 & 1.09 & 16.11 & 0.91 & 1.00 \\
\hline
\multicolumn{12}{c}{NGC 2482} \\
\hline
1 & 5698593124091694720 & 118.8078 & -24.2677 & 12.80 & 0.35 & -4.72 & 2.29 & 0.75 & --- & 0.94 & 1.00 \\
2 & 5698587244265214208 & 118.7942 & -24.2728 & 11.12 & 0.14 & -4.67 & 2.21 & 0.77 & 33.03 & 1.27 & 1.00 \\
3 & 5698587244273305984 & 118.7900 & -24.2782 & 16.66 & 1.17 & -4.67 & 2.29 & 0.63 & --- & 1.07 & 0.99 \\
\ldots & \ldots & \ldots & \ldots & \ldots & \ldots & \ldots & \ldots & \ldots & \ldots & \ldots & \ldots \\
116 & 5698513821813592448 & 119.0220 & -24.7615 & 15.94 & 0.97 & -4.71 & 2.24 & 0.78 & --- & 0.99 & 0.96 \\
\hline
\bottomrule
\end{tabular}
\end{table*}

\subsection{Inner Kinematics and Velocity Ellipsoid Parameters}

The internal kinematic properties of the program stars -- i.e., the probable cluster members with available {\it Gaia} DR3 radial velocities used in Section~\ref{sec:mean_rv} (188 stars for NGC 2423 and 29 stars for NGC 2482) -- were investigated using their three-dimensional space velocities derived from $Gaia$ astrometry and available radial velocity measurements. Stellar positions and velocities were initially expressed in the equatorial reference frame, which is naturally suited to the observed proper motions (PMs) and trigonometric parallaxes. The explicit transformation from angular observables to Cartesian equatorial velocity components is standard and is therefore summarized only briefly here.

The equatorial velocity components $(V_x,~V_y,~V_z$; km~s$^{-1}$) were computed by combining the line-of-sight velocity with the transverse motions inferred from proper motions and distances. This formulation allows the radial and tangential contributions to be treated consistently and enables the construction of velocity components that depend exclusively on either radial velocities or proper motions when required \citep{Smart68}.

The transformation from equatorial velocities to Galactic space velocities $(U,~V,~W;$ km~s$^{-1}$) was performed using an orthogonal rotation matrix defined by the adopted equatorial coordinates of the North Galactic Pole and the zero point of Galactic longitude. We followed the formalism of \citet{1987AJ.....93..864J}, while also verifying the results using the high-precision equatorial--Galactic transformation matrix based on the SPECFIND v2.0 catalog \citep{2011AA...536A.102L}. Both approaches yield consistent velocity components within the numerical uncertainties, and the latter was adopted for the final analysis due to its higher numerical accuracy. Details of the transformation matrices are given in \citet{Johnson_1987}.

To characterize the internal kinematics of the system, we quantified the velocity dispersion tensor constructed from the Galactic velocity components. The second-order central moments of the velocity distribution were computed to form the symmetric covariance matrix, whose eigenvalues correspond to the squared principal velocity dispersions. Following the methodology described by \citet{elsanhoury24, elsanhouryAN}, the principal dispersions $(\sigma_1,~\sigma_2,~\sigma_3;$ km~s$^{-1}$) were obtained by solving the characteristic equation of the covariance matrix \citep{elsanhoury24, elsanhouryAN}.

The resulting Velocity Ellipsoid Parameters (VEPs) provide a compact and physically meaningful description of the internal dynamical state of the system, enabling a direct comparison with previous studies of open clusters and moving groups. Figure~\ref{Fig:13} shows the distribution of the Galactic velocity components $(U,~V,~W)$ for the clusters, illustrating their kinematic coherence and relative dispersion.

\subsection{CP}

The convergent point (CP) method is used to characterize the bulk motion of an open cluster by exploiting the common space motion of its member stars. Due to perspective effects, the proper motion vectors of cluster members appear to converge toward a single point on the celestial sphere, referred to as the convergent point. When combined with radial velocity information, this method provides a reliable framework for deriving the three-dimensional kinematics of star clusters \citep{Chupina2001,Chupina2006}.

In this study, we applied the apex-diagram (AD-diagram) formulation of the CP method to the open clusters NGC 2423 and NGC 2482. To avoid introducing additional noise caused by coordinate transformations of individual measurements with differing uncertainties, we derived the bulk space velocity components by directly transforming the reliable mean observational parameters of the clusters ($\overline{\alpha},~\overline{\delta},~\overline{d},~\overline{\mu_\alpha \cos\delta},~\overline{\mu_\delta},~\overline{V_{\rm r}}$). The mean Cartesian equatorial space velocity components were derived from the mean cluster parameters using the standard formalism described in \citet{Melchior1958}.

The equatorial coordinates of the cluster apex $(A_0,~D_0)$ are then derived from these mean space velocity components $(\overline{V}_x,~\overline{V}_y,~\overline{V}_z)$ as
\begin{equation}
A_0 = \tan^{-1}\left(\frac{\overline{V}_y}{\overline{V}_x}\right),
\end{equation}
\begin{equation}
D_0 = \tan^{-1}\left(\frac{\overline{V}_z}{\sqrt{\overline{V}_x^{\,2}+\overline{V}_y^{\,2}}}\right).
\end{equation}

\begin{figure*}[t]
\centering
\includegraphics[width=0.35\linewidth]{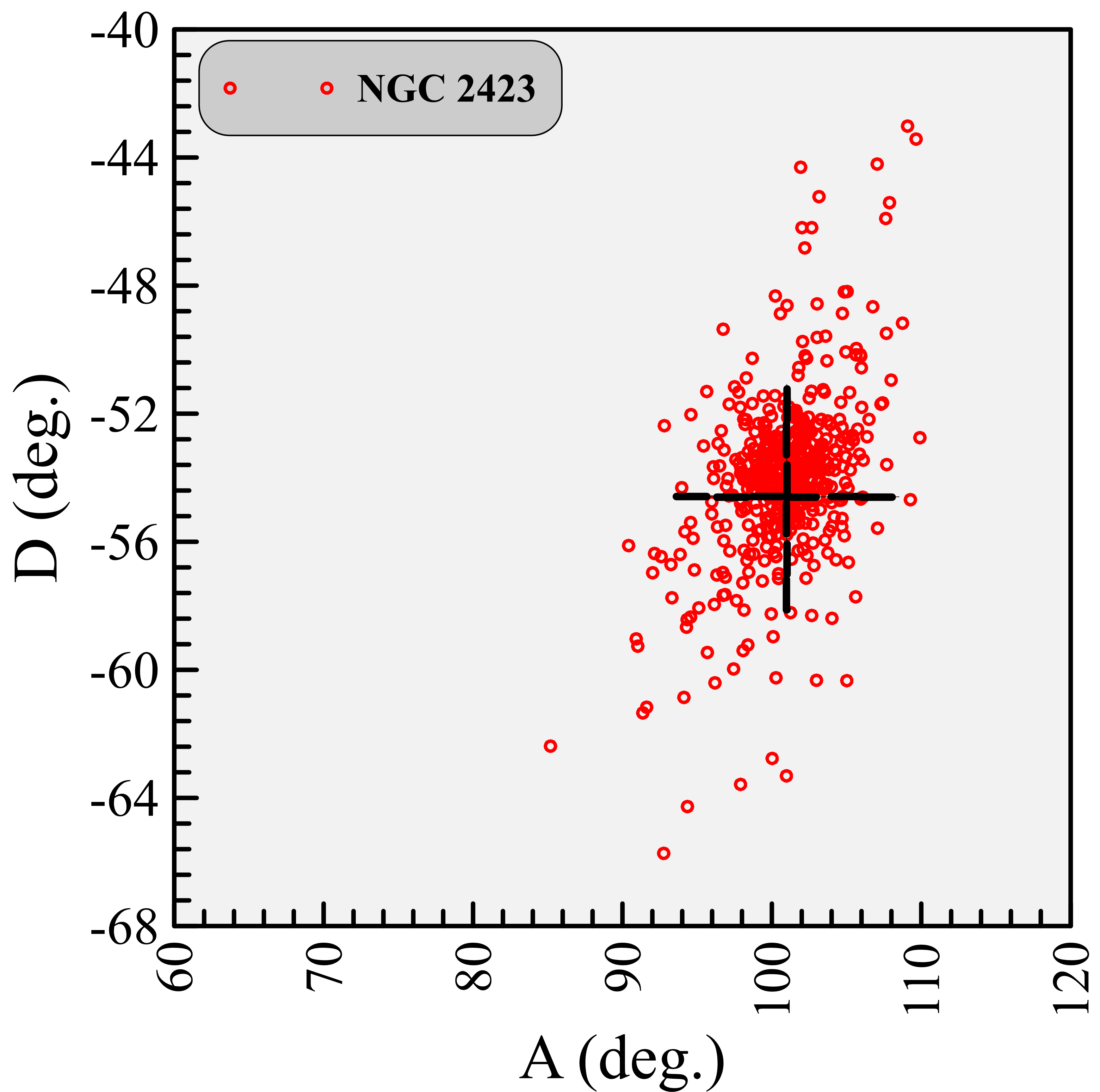}
\includegraphics[width=0.35\linewidth]{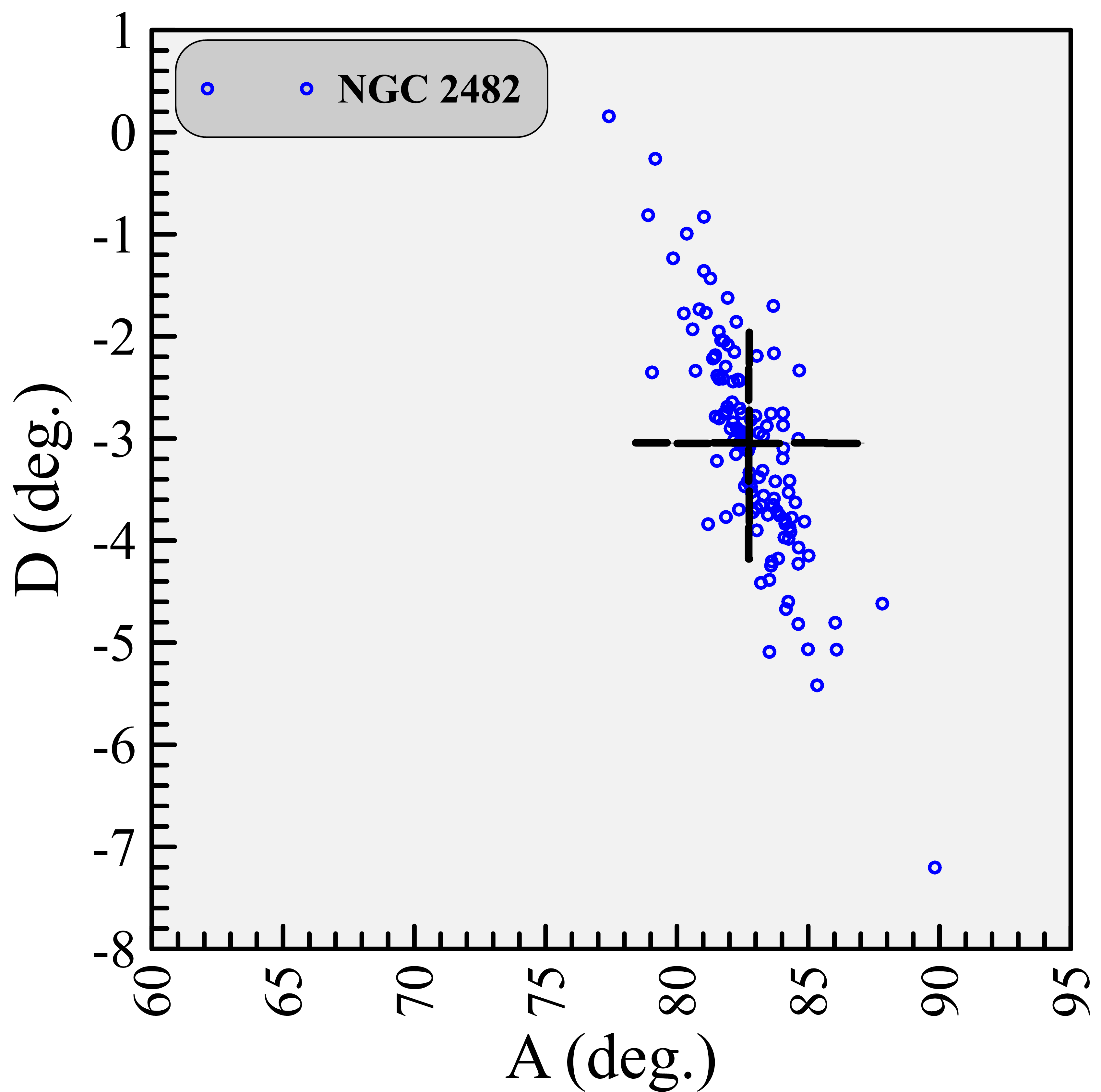}
\caption{Apex diagrams (AD-diagrams) for NGC 2423 (left) and NGC 2482 (right), showing the distribution of stellar apex coordinates $(A_0,~D_0)$ for cluster members. Crosses indicate the mean apex positions, with error bars representing the standard deviations in $A_0$ and $D_0$.}
\label{Fig:12}
\end{figure*}

Although the bulk kinematics are derived from the mean parameters, we also computed individual stellar apex coordinates to assess kinematic coherence. Figure~\ref{Fig:12} shows the AD-diagrams for NGC 2423 and NGC 2482. In both clusters, the apex coordinates of individual member stars form compact distributions, indicating that the stars share a common velocity vector. The mean apex positions, marked by cross symbols, are consistent within uncertainties.

\begin{figure*}
\centering
\includegraphics[width=0.45\linewidth]{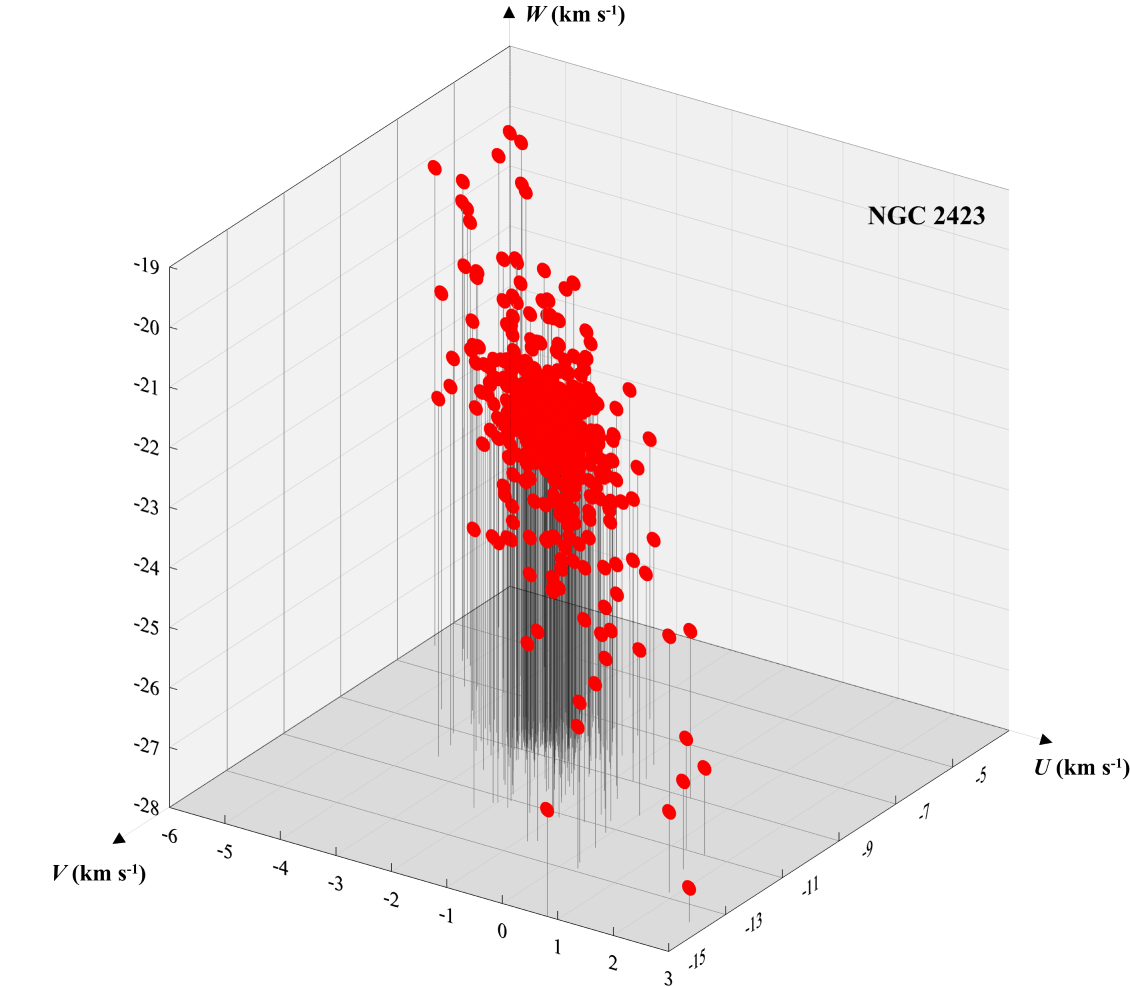}
\includegraphics[width=0.45\linewidth]{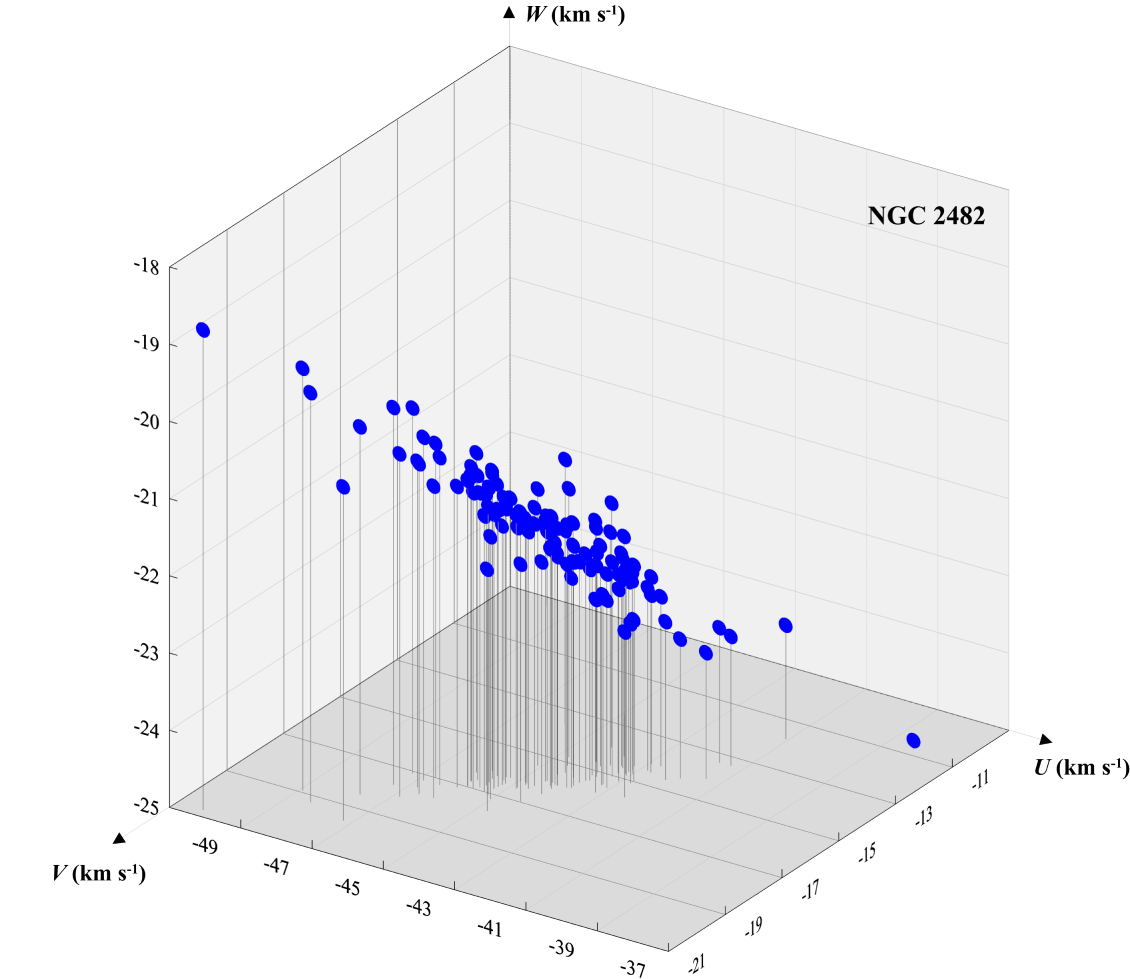}
\caption{Galactic space velocity components $(U,~V,~W)$ of member stars in NGC 2423 (left) and NGC 2482 (right).}
\label{Fig:13}
\end{figure*}

The transformation from the mean equatorial velocity components $(\overline{V}_x,~\overline{V}_y,~\overline{V}_z)$ to the Galactic velocity components $(\overline{U},~\overline{V},~\overline{W})$ was performed using the standard approach outlined in \citet{Cinar2026}. For illustrative purposes, Figure~\ref{Fig:13} presents the distributions of the individual Galactic velocity components for NGC 2423 and NGC 2482. The two clusters exhibit closely overlapping velocity distributions and similar mean space velocities, consistent with a common bulk motion.

\subsection{Galactic Space Velocities and Solar Motion}

By reciprocity, the mean Galactic space velocity of a cluster relative to the Sun is numerically equal and opposite to the apparent velocity of the Sun relative to that cluster's rest frame. We exploit this reciprocity as an independent, purely kinematic consistency check on our derived cluster velocities: the resulting Solar apex position and speed, computed relative to each cluster taken individually, can be compared against the well-established Solar motion relative to the Local Standard of Rest (LSR; $S_\odot\approx13$--20~km~s$^{-1}$ toward $(l_A,b_A)\approx(56^\circ,23^\circ)$ in the classical determinations). This comparison, following the same procedure applied in our previous cluster kinematic studies \citep{Elsanhoury2016,Elsanhoury2022}, serves to verify that the sign conventions, coordinate transformations, and mean space velocities adopted for NGC 2423 and NGC 2482 are internally consistent, rather than to derive a new measurement of the Solar motion itself. In this context, the Sun’s velocity components and apex position are derived relative to the bulk motion of the analyzed clusters, assuming that the cluster members define a local kinematic reference frame.

For NGC 2423 and NGC 2482, the mean Galactic velocity components $(\overline{U}, \overline{V}, \overline{W})$ were obtained directly from the transformation of the mean cluster parameters as described in the previous section. The corresponding Solar velocity components $(U_{\odot},~V_{\odot},~W_{\odot})$ are then computed following \citet{Elsanhoury2016,Elsanhoury2022} as
\begin{equation}
\begin{split}
U_\odot&=-\overline{U} = -S_\odot\cos{l_A} \cos{b_A},\\
V_\odot&=-\overline{V} = -S_\odot\cos{l_A} \sin{b_A},\\
W_\odot&=-\overline{U} = -S_\odot\sin{b_A}.
\end{split}
\end{equation}

The magnitude of the Solar velocity relative to the cluster reference frame is given by
\begin{equation}
\label{Eq:SolarSpeed}
S_{\odot} = \sqrt{\overline{U}^{\,2} + \overline{V}^{\,2} + \overline{W}^{\,2}} =\sqrt{\left(X_{\odot}^{\bullet}\right)^2+\left(Y_{\odot}^{\bullet}\right)^2+\left(Z_{\odot}^{\bullet}\right)^2}.
\end{equation}

The Sun's velocities were relative to these groups and the same axes are defined as ($X_{\odot}^{\bullet} = -\overline{V_x}$), ($Y_{\odot}^{\bullet} = -\overline{V_y}$), and ($Z_{\odot}^{\bullet} = -\overline{V_z}$), taking into account the position along the $x,~y,$ and $z$ - axes in the coordinate system centered around the Sun (i.e., equatorial coordinates). On the other hand, $l_{\rm A}$ and $b_{\rm A}$ denote the Galactic longitude and latitude of the Solar apex, respectively. 

Using the mean space velocities of NGC 2423 and NGC 2482, we estimated the Solar velocity magnitude. The resulting values are summarized in Table~\ref{tab:dynkin}. We also derived the corresponding Solar velocity components, including the radial velocities.

The Galactic coordinates of the Solar apex $(l_{\rm A},~b_{\rm A})~\&~(\alpha_{\rm A},~\delta_{\rm A})$ are determined from
\begin{equation}
\label{Eq:SolarApex}
l_{\rm A} = \tan^{-1}\left(\frac{-\overline{V}}{\overline{U}}\right), \qquad
b_{\rm A} = \sin^{-1}\left(\frac{-\overline{W}}{S_{\odot}}\right),
\end{equation}
\centerline{$\&$}
\begin{equation}
\alpha_{A} = \tan^{-1}\Bigg(\frac{Y_{\odot}^{\bullet}}{X_{\odot}^{\bullet}}\Bigg),
\qquad
\delta_{A} = \tan^{-1}\Bigg(\frac{Z_{\odot}^{\bullet}}{\sqrt{\left(X_{\odot}^{\bullet}\right)^2+\left(Y_{\odot}^{\bullet}\right)^2}}\Bigg).
\end{equation}

\section{Luminosity and Mass Function}

\begin{figure}[!ht]
\centering
\includegraphics[width=0.8\linewidth]{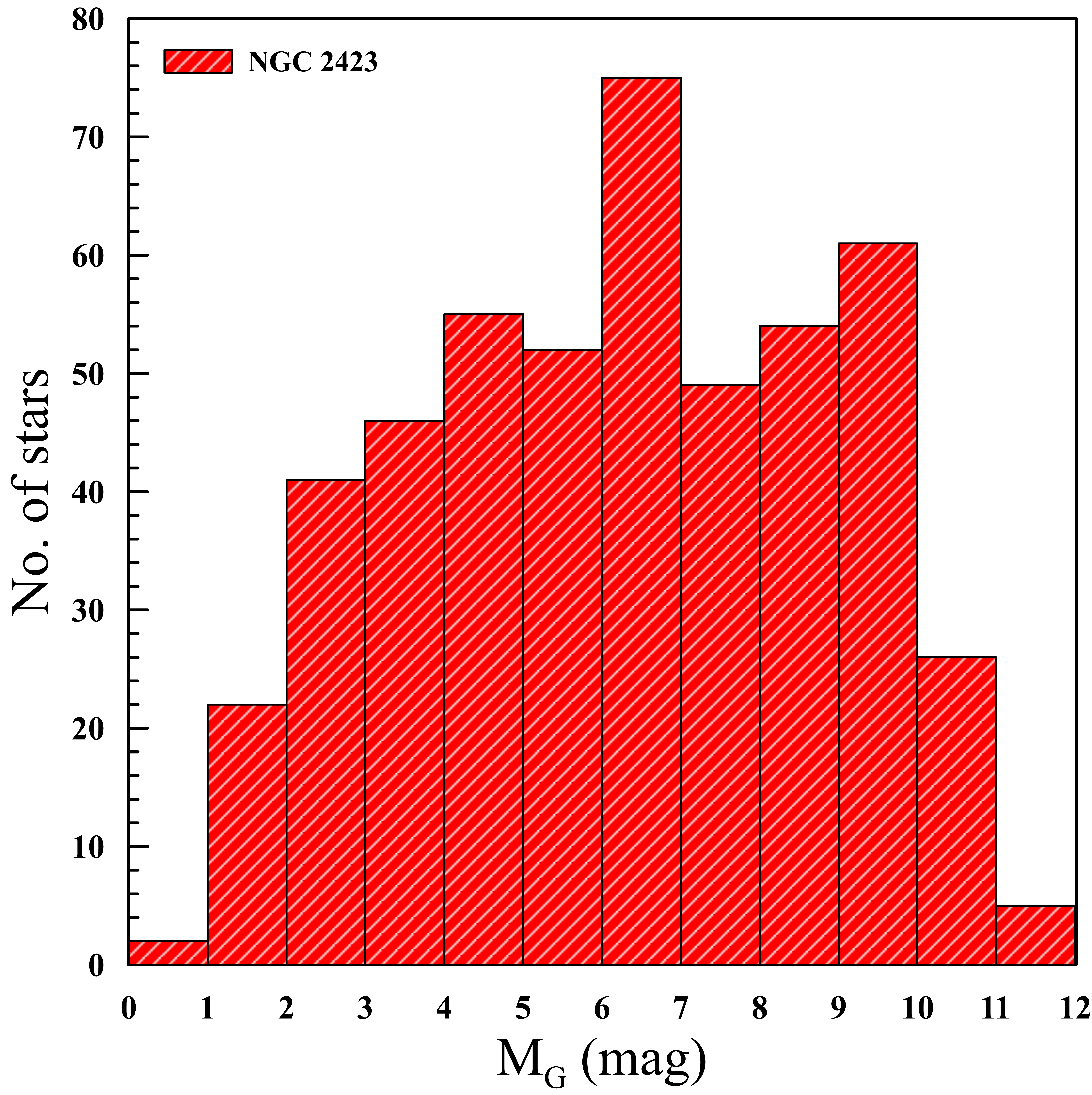}\\
\includegraphics[width=0.8\linewidth]{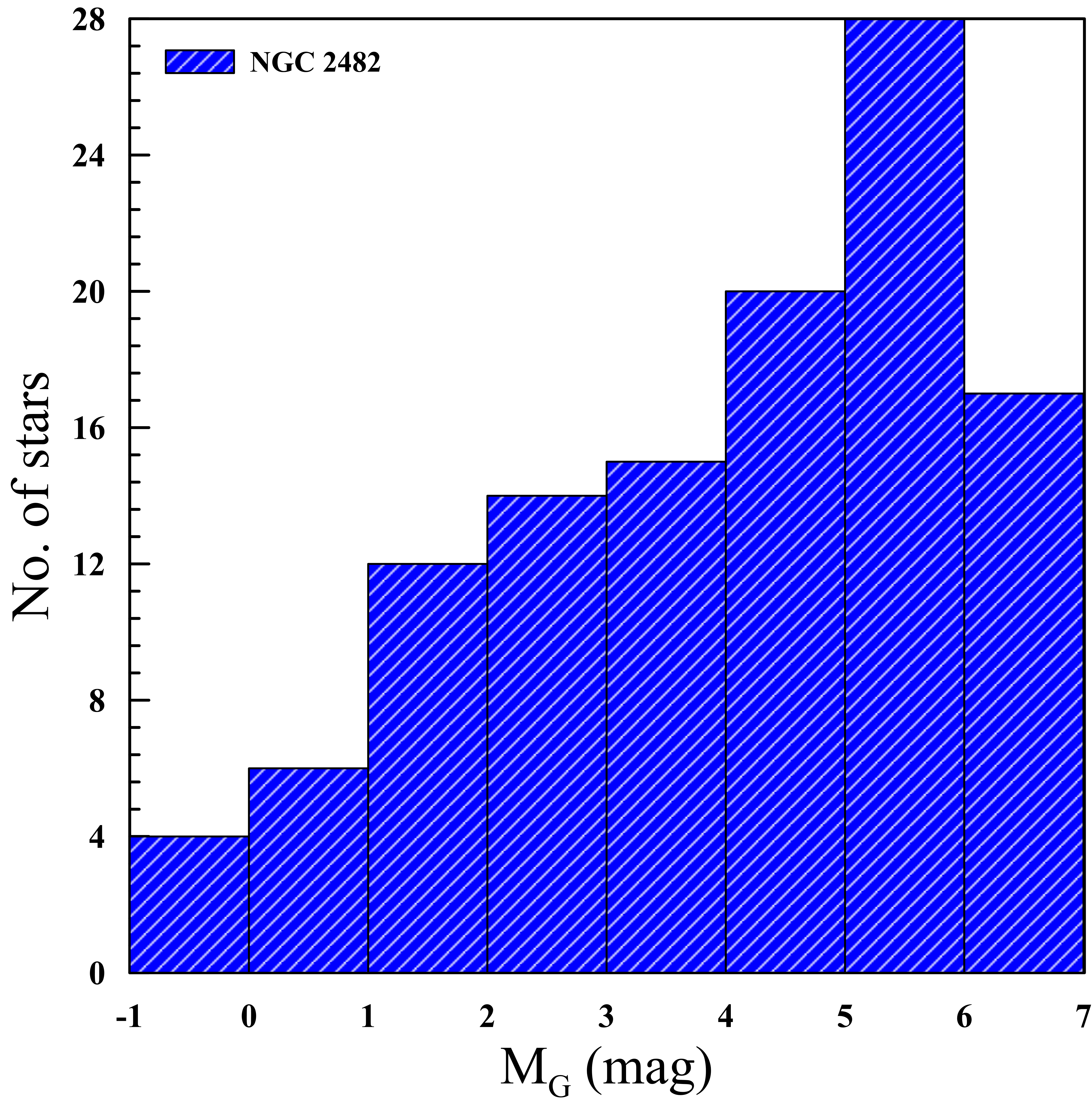}
\caption{Luminosity functions of NGC 2423 and NGC 2482 in the $G$ absolute band.}
\label{Fig:6}
\end{figure}

The luminosity function (LF) describes the distribution of cluster members as a function of absolute magnitude and provides insight into the stellar content of the system \citep{Haroon2017}. For each cluster, probable members were selected based on the membership analysis described earlier. \textcolor{black}{Apparent $G$-band magnitudes were converted into absolute magnitudes using the adopted observed distance moduli, $(m-M)_{\rm obs}$, which already account for the G-band extinction.} The resulting absolute magnitudes were grouped into bins to construct the LF, ensuring adequate star counts in each interval. The $G$-band luminosity functions for NGC 2423 and NGC 2482 are shown in Figure~\ref{Fig:6}. The mean absolute magnitudes were found to be $\overline{M_G}$ = 6.285 mag for NGC 2423 and $\overline{M_G}$ = 3.986 mag for NGC 2482.

The mass function (MF) reflects the present-day distribution of stellar masses within a cluster and can be derived from the LF using an appropriate mass--luminosity relation. Assuming a power-law form, the MF is expressed as \citep{Salpeter1955}
\begin{equation} \label{Eq:MF}
\frac{dN}{dM} \propto M^{-\alpha},
\end{equation}
where $dN/dM$ is the number of stars per unit mass interval and $\alpha$ is the MF slope. Stellar masses were estimated using a polynomial mass--luminosity relation obtained from the best-fitting theoretical isochrones:
\begin{equation} \label{Eq:MLR}
M_{\rm c} = a_0 + a_1 M_G + a_2 M_G^2 + a_3 M_G^3 + a_4 M_G^4,
\end{equation}
where the coefficients $a_0$--$a_4$ were determined separately for each cluster.

\begin{figure}[!ht]
\centering
\includegraphics[width=0.9\linewidth]{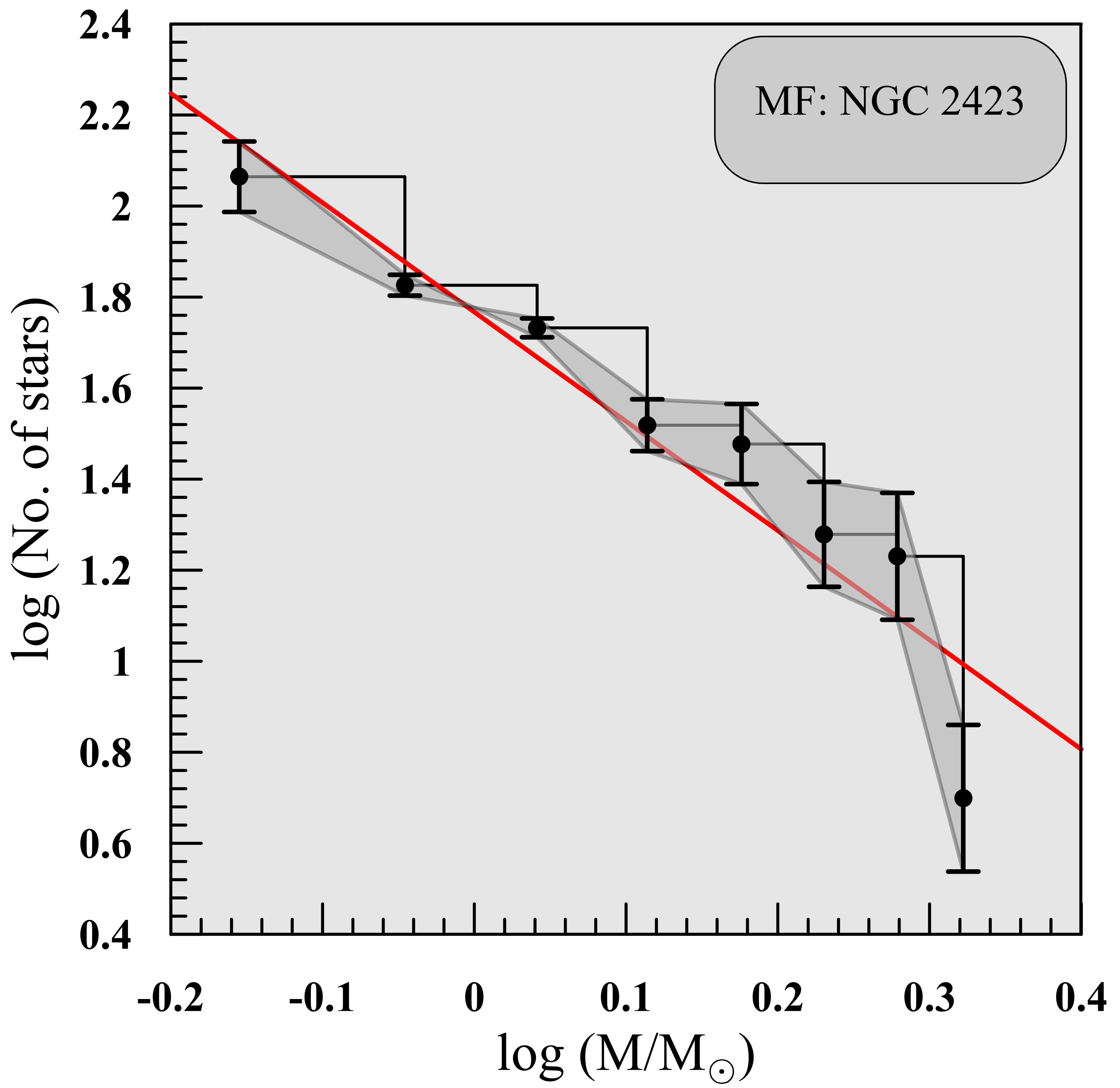}\\
\includegraphics[width=0.9\linewidth]{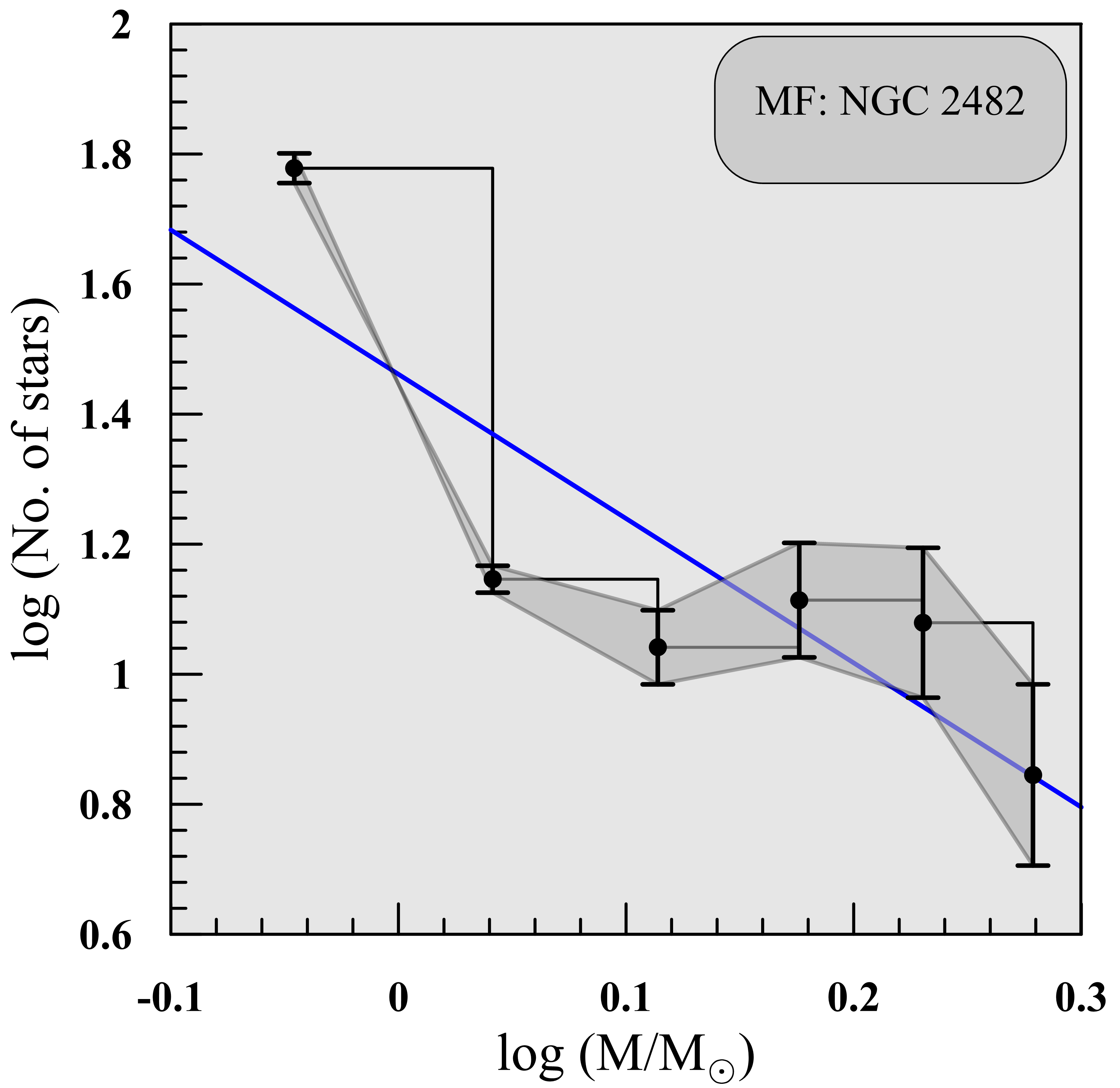}
\caption{Mass functions of NGC 2423 and NGC 2482 fitted with a power-law.}
\label{Fig:7}
\end{figure}

From the derived mass distributions, the total cluster mass ($M_{\rm C}$), mean stellar mass ($\overline{M_{\rm C}}$), and MF slope ($\alpha$) were calculated. The resulting parameters for NGC 2423 and NGC 2482 are listed in Table~\ref{Tab:5}. The logarithmic mass functions, along with the fitted power-law slopes, are shown in Figure~\ref{Fig:7}. The power-law fit for NGC 2423, based on 488 member stars, is well constrained and closely tracks the binned data over the full mass range probed. For NGC 2482, the fit is visibly less tight, particularly at $\log(M/M_\odot)\lesssim0.05$, where the binned points show larger scatter around the fitted line; this reflects the smaller final membership list of NGC 2482 (116 stars) and the correspondingly larger Poisson uncertainties in the low-number mass bins used to construct the MF, rather than a fundamentally different functional form. \textcolor{black}{The fitted MF slopes are $\alpha=2.41\pm0.06$ for NGC 2423 and
$\alpha=2.22\pm0.03$ for NGC 2482.} We therefore regard the derived slope $\alpha=2.22\pm0.03$ for NGC 2482 as more uncertain in a systematic (not merely statistical) sense than the formal fit uncertainty suggests, and recommend a larger member sample -- e.g., extending the survey to fainter magnitudes or larger radii, informed by the discussion of Figure~\ref{fig:rdps} above -- before over-interpreting small differences in $\alpha$ between the two clusters. While some dynamically complex open clusters are reported to exhibit multimodal mass functions (e.g., \citealt{AHMED2026}), the MFs of both NGC 2423 and NGC 2482 are adequately, if not perfectly, described by a standard power-law distribution.

\begin{table}[t]
\centering
\caption{Luminosity and mass function parameters of the studied clusters.}
\begin{tabular}{lcc}
\hline
\textcolor{black}{Parameter} & \textcolor{black}{NGC 2423} & \textcolor{black}{NGC 2482} \\
\hline
\hline
$\overline{M_G}$ (mag) & 6.285 & 3.986 \\
$M_{\rm C}$ ($M_{\odot}$) & 436 $\pm$ 21 & 142 $\pm$ 12 \\
$\overline{M_{\rm C}}$ ($M_{\odot}$) & 0.894 & 1.220 \\
$a_0$ & 2.710 $\pm$ 0.60 & 2.057 $\pm$ 0.07 \\
$a_1$ & -0.543 $\pm$ 0.03 & -0.027 $\pm$ 0.02 \\
$a_2$ & 0.044 $\pm$ 0.01 & -0.115 $\pm$ 0.01 \\
$a_3$ & 0.001 $\pm$ 0.00 & 0.021 $\pm$ 0.00 \\
$a_4$ & 0.001 $\pm$ 0.00 & -0.001 $\pm$ 0.00 \\
$\alpha$ & 2.41 $\pm$ 0.06 & 2.22 $\pm$ 0.03 \\
\hline
\end{tabular}
\label{Tab:5}
\end{table}

\begin{figure*}[htb!]
\includegraphics[width=0.99\linewidth]{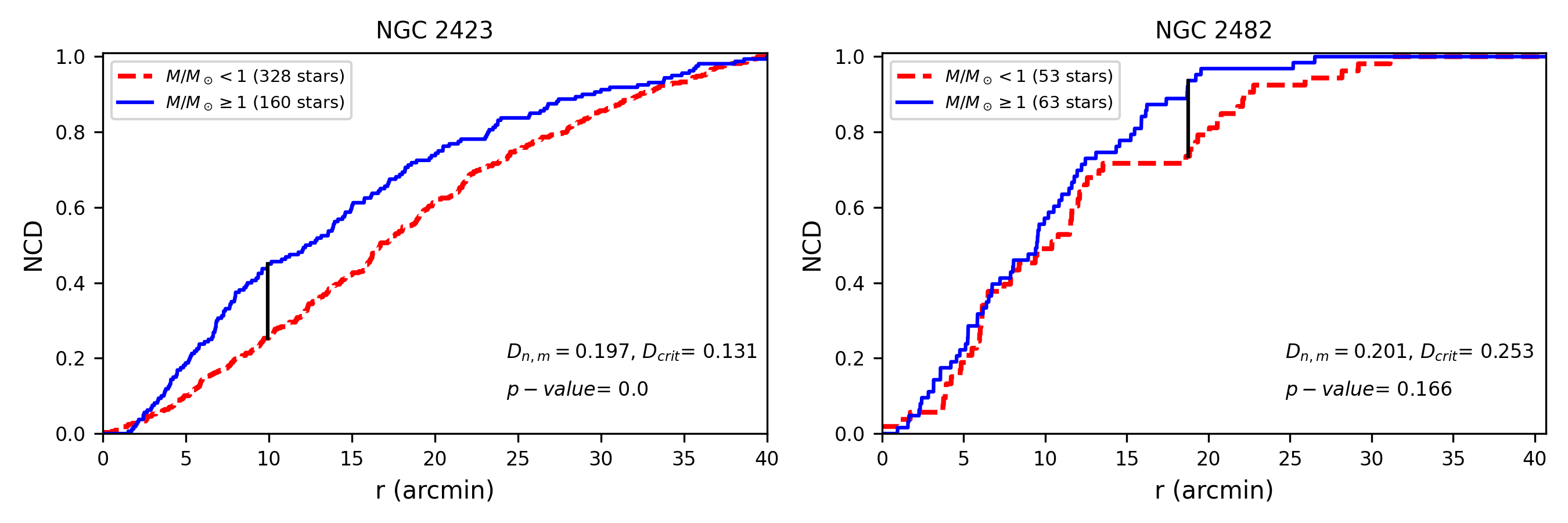}
\caption{Normalized cumulative distributions (NCD) of the radial positions of high-mass ($M\geq1\,M_\odot$) and low-mass ($M<1\,M_\odot$) members, used in the Kolmogorov--Smirnov test for mass segregation in the investigated clusters.
\label{KS_tests}}
\end{figure*}

\subsection{Evidence for Mass Segregation}

Mass segregation refers to the preferential concentration of higher-mass stars toward the central regions of a star cluster, while lower-mass stars are more widely distributed. This effect may arise from the cluster’s initial formation conditions or develop over time through two-body relaxation and dynamical evolution. To investigate the presence of mass segregation in NGC 2423 and NGC 2482, we examined the spatial distributions of stars in different mass ranges.

Cluster members were divided into high-mass and low-mass subsamples using a mass threshold of $M = 1\,M_{\odot}$. The cumulative radial distributions of these subsamples -- hereafter denoted NCD (normalized cumulative distribution), and plotted as such in Figure~\ref{KS_tests} -- were constructed as a function of distance from the cluster center, with each distribution normalized to unity at the maximum radius so that the two subsamples (of generally different size) can be compared directly.
In this work, we used the Kolmogorov–Smirnov statistical test to look for signs of mass segregation in the investigated clusters. The KS-test searches for the maximum difference between the cumulative radial distributions of two samples with sizes n and m, $D_{n,m}$ computed using the following relation:
\begin{equation}
D_{n,m} = sup_x |F(x)-G(x)|
\end{equation}
The null hypothesis, which is that the two distributions are similar, is rejected if $D_{n,m}$ is greater than the critical value $D_{crit}$ calculated for the same sizes as those of the two samples and for a selected level of significance $\alpha$.

Figure \ref{KS_tests} shows the cumulative radial distributions of members of the two clusters in two mass ranges for each cluster, where the difference between the distribution and the probability of resemblance, p-value, is shown in Figure \ref{KS_tests}. 
The figure shows that in the NGC 2423 OC, higher-mass stars are found to be more centrally concentrated compared to their lower-mass counterparts, suggesting the presence of mass segregation for $\alpha$ = 0.05 (or 5\% resemblance). On the other hand, there is no significant difference between the distributions of the low mass and high mass stars in NGC 2482 suggesting the absence of clear mass segreagation in this cluster.

\subsection{Galactic Orbital Properties}

The Galactic orbital dynamics of NGC 2423 and NGC 2482 were examined through numerical orbit integrations in a static, three-component Galactic potential. We adopted the \texttt{MWPotential2014} model, which consists of a bulge described by a power-law density profile with an exponential cutoff, a Miyamoto--Nagai disc, and a Navarro--Frenk--White dark matter halo \citep{Bovy_2012}, as implemented in the \texttt{galpy} Python package \citep{Bovy_2015}. The orbit integrations were initialized using the present-day six-dimensional phase-space information of the clusters $(\alpha,~\delta,~d,~\mu_{\alpha}\cos\delta,~\mu_{\delta},~V_{\rm r})$ derived from our observational data.

\begin{figure*}[!ht]
    \centering
    \includegraphics[width=0.48\linewidth]{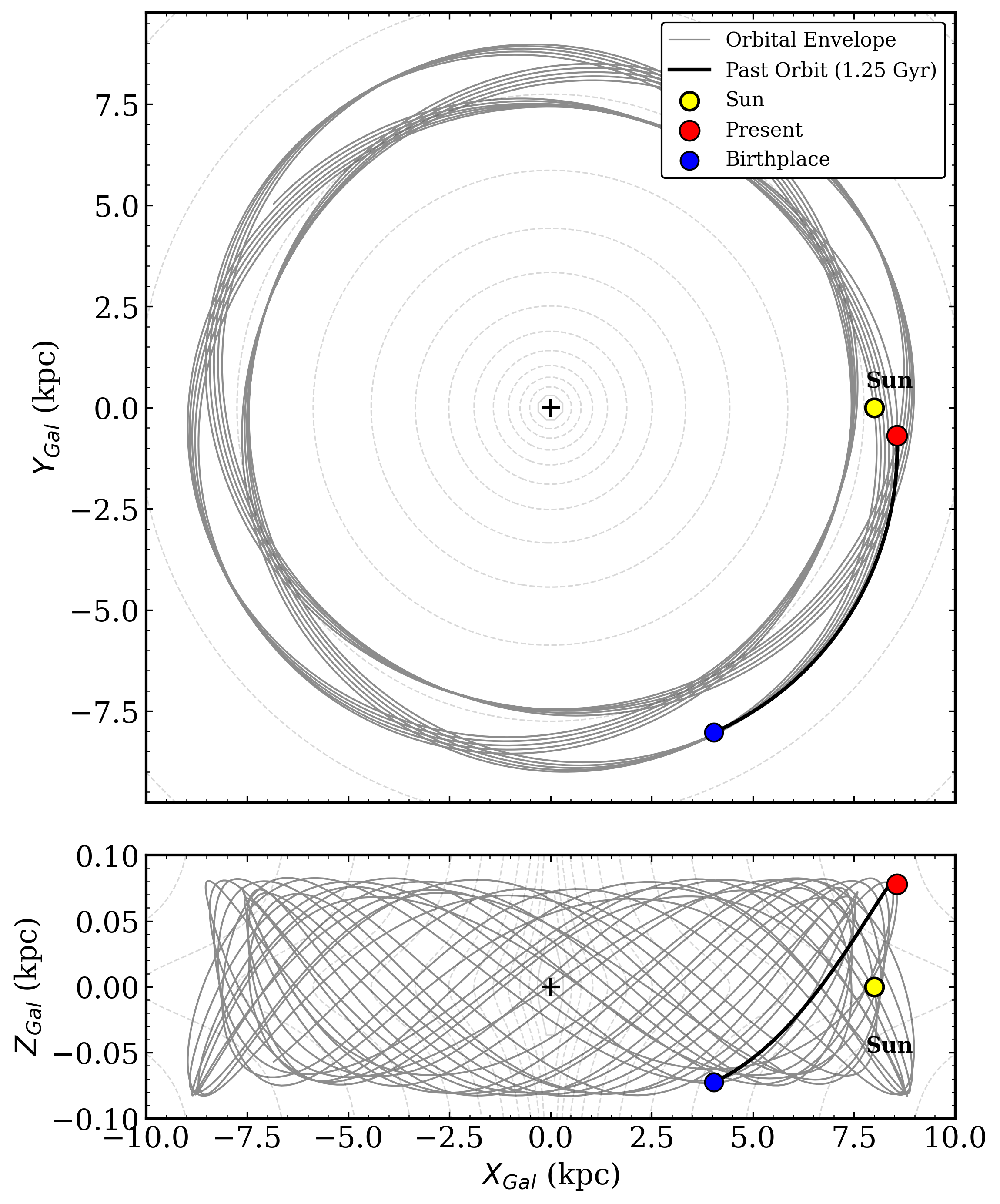}
    \includegraphics[width=0.48\linewidth]{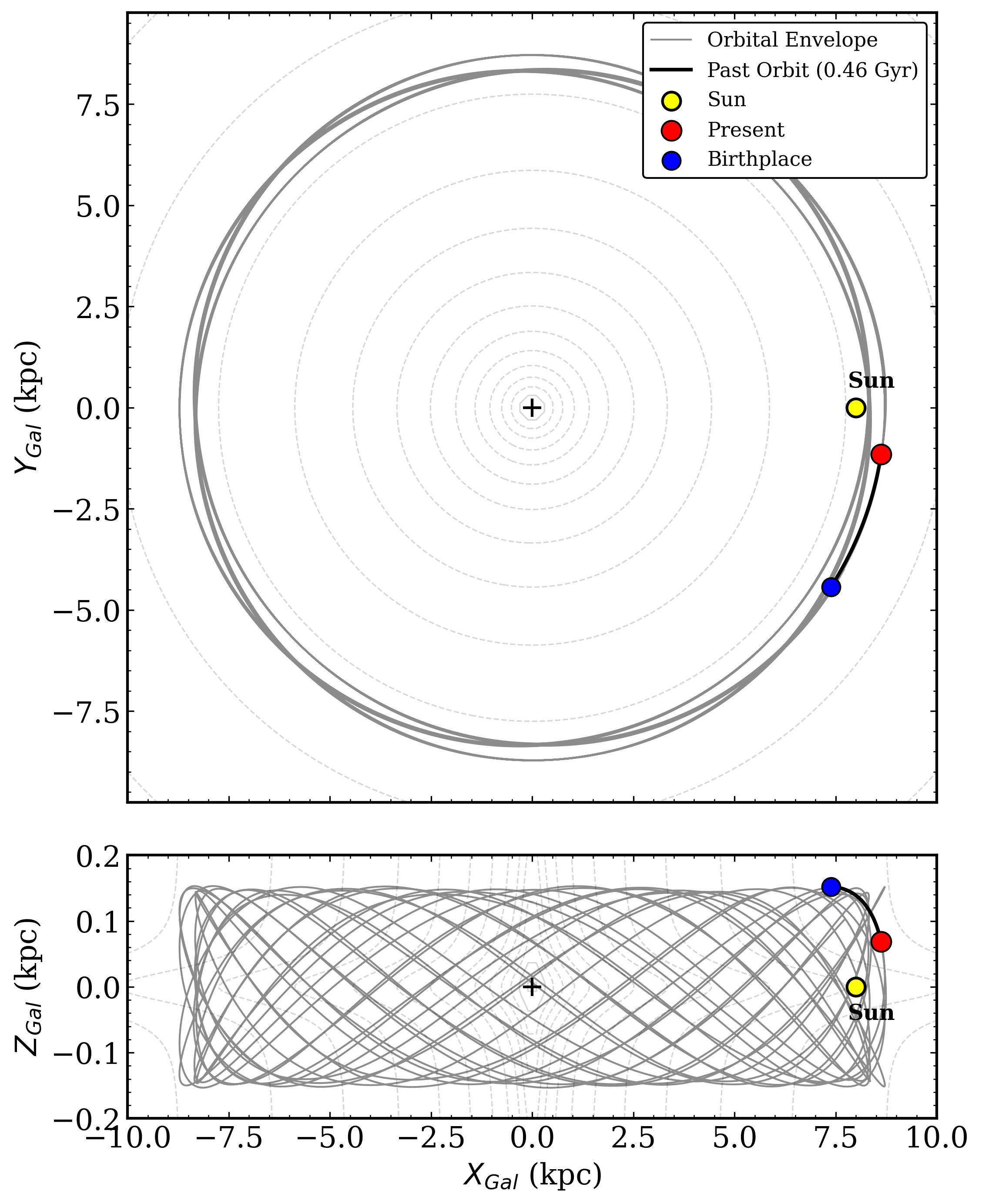}
    \caption{Orbital evolution of the open clusters NGC 2423 (left) and NGC 2482 (right) in a Galactic potential. The top panels show the orbital projections in the Galactic $X$--$Y$ plane, while the bottom panels show the $X$--$Z$ projections. The grey shaded regions indicate the orbital envelopes tracing the region of phase space occupied by the clusters. The solid black curves represent backward orbit integrations over the cluster ages (1.25~Gyr for NGC 2423 and 0.46~Gyr for NGC 2482). The present-day positions of the Sun and the clusters are marked by yellow and red circles, respectively, while blue circles indicate the inferred birth positions. The grey background contours show the logarithmic density distribution of the \texttt{MWPotential2014} model.}
    \label{fig:orbits}
\end{figure*}

The orbits were integrated forward over several orbital periods to assess their stability and to characterize the overall orbital structure \citep{Alzahrani2025a, Alzahrani2025b, Bisht2026}. From the resulting trajectories, we derived the principal orbital parameters, including the apogalactic distance ($R_{\rm a}$), perigalactic distance ($R_{\rm p}$), and the maximum vertical distance from the Galactic plane ($Z_{\rm max}$). We additionally computed the radial action ($J_R$), which quantifies the amplitude of radial oscillations and provides a dynamical measure of orbital eccentricity, as well as the vertical component of the angular momentum ($L_z$), which traces the rotational support and sense of motion with respect to the Galactic disk. The orbital eccentricity was computed as $e = (R_{\rm a} - R_{\rm p}) / (R_{\rm a} + R_{\rm p})$, while the orbital period ($T_{\rm p}$) was estimated from the azimuthal motion along the orbit. The resulting orbital configurations are illustrated in Figure~\ref{fig:orbits}.

To obtain approximate constraints on the past radial locations of the clusters, backward orbit integrations were performed over their respective ages, from which the corresponding birthplace radii ($R_{\rm birth}$) were inferred. The uncertainties associated with the derived orbital parameters were estimated using a simplified error propagation approach that accounts for the observational uncertainties in distance and proper-motion measurements. The resulting orbital properties, summarized in Table~\ref{tab:dynkin}, indicate that both clusters follow low-eccentricity orbits ($e < 0.1$) and remain confined close to the Galactic plane ($|Z_{\rm max}| < 0.2$~kpc), consistent with regular disc-like orbital behavior.

\begin{table*}[!ht]
\centering
\footnotesize
\caption{Derived kinematic and dynamical parameters of the open clusters NGC 2423 and NGC 2482.}
\label{tab:dynkin}
\renewcommand{\arraystretch}{1}
\setlength{\tabcolsep}{6pt}
\begin{tabular}{lcc}
\hline
\hline
Parameter & NGC 2423 & NGC 2482 \\
\hline
\multicolumn{3}{c}{Kinematical and VEPs} \\
\hline
($\overline{V_x},~\overline{V_y}, ~\overline{V_z}$; km\,s$^{-1}$) & -2.76 $\pm$ 0.06, 14.27 $\pm$ 3.78, -20.10 $\pm$ 4.48 & 6.44 $\pm$ 0.40, 51.00 $\pm$ 7.14, -2.74 $\pm$ 0.06  \\
($\overline{U},~\overline{V}, ~\overline{W}$; km\,s$^{-1}$) & -2.53 $\pm$ 0.06,~-22.82 $\pm$ 4.78,~-9.37 $\pm$ 3.06 & -43.49 $\pm$ 6.60, -21.77 $\pm$ 4.67, -16.88 $\pm$ 4.11  \\
($l_1,~m_1,~n_1$)$^o$ & 0.100, 0.920, 0.379 & 0.845, 0.422, 0.328 \\
($l_2,~m_2,~n_2$)$^o$ & -0.783, -0.162, 0.601  & -0.226, 0.838, -0.496  \\
($l_3,~m_3,~n_3$)$^o$ & -0.783, -0.162, 0.601  & -0.485, -0.345, -0.804 \\
($\sigma_1,~\sigma_2,~\sigma_3$; km s$^{-1}$) & 24.80 $\pm$ 4.98, 1.27 $\pm$ 0.08, 0.66 $\pm$ 0.01 & 51.28 $\pm$ 7.16, 1.73 $\pm$ 0.07, 0.49 $\pm$ 0.01  \\
$\sigma_o$ (km\ s$^{-1}$)  & 24.84 $\pm$ 4.98 & 51.31 $\pm$ 7.16 \\

$S_{\odot}$ (km\ s$^{-1}$)      & $24.80\pm4.80$ & $51.48\pm7.17$ \\
\hline
\multicolumn{3}{c}{Dynamical Parameters} \\
\hline
$A_0^{(o)}$           & $100.92\pm0.10$ & $82.81\pm0.11$ \\
$D_0^{(o)}$           & $-54.12\pm0.13$& $-3.05\pm0.06$ \\
$l_A^{(o)}$             & $-83.67$         & $-26.60$         \\
$b_A^{(o)}$             & $22.19$         & $19.14$          \\
$\alpha_A^{(o)}$                   &-79.08           & 82.80             \\
$\delta_A^{(o)}$                   & 54.12           & 3.05             \\
$Z_{\rm max}$ (kpc)          & $0.084\pm0.003$& $0.153\pm0.018$\\
$R_{\rm a}$ (kpc)            & $9.044\pm0.062$& $8.719\pm0.088$\\
$R_{\rm p}$ (kpc)            & $7.442\pm0.027$& $8.209\pm0.239$\\
$R_{\rm m}$ (kpc)            & $8.243\pm0.034$& $8.464\pm0.128$\\
$L_z$ (kpc km s$^{-1}$)      & $1793.80\pm1.88$ & $1848.31\pm32.13$ \\
$J_R$ (kpc km s$^{-1}$)        & $11.55\pm0.97$   & $1.17\pm1.16$ \\
$e$                          & $0.097\pm0.005$& $0.030\pm0.009$\\
$T_{\rm p}$ (Myr)            & $231\pm1$              & $238\pm5$              \\
$R_{\rm Birth}$ (kpc)        & 9.009 $\pm$ 0.086 & $8.282\pm0.161$\\

\hline
\hline
\end{tabular}
\end{table*}

\subsection{Internal Dynamical Evolution}\label{sec:internal_dyn}

The internal dynamical evolution of an open cluster is driven by gravitational interactions among its member stars, leading to processes such as energy exchange through stellar encounters and the gradual redistribution of stellar orbits. These interactions cause the system to evolve toward a dynamically relaxed configuration over a characteristic timescale known as the dynamical relaxation time \citep{Lamers2006}. This timescale provides a quantitative measure of whether a cluster has preserved its initial dynamical structure or has reached equilibrium.

The dynamical relaxation time is estimated following \citet{Spitzer1971} as
\begin{equation} \label{Eq:relax}
T_{\rm relax} = \frac{8.9 \times 10^{5}\, N^{1/2} \, R_{\rm h}^{3/2}}
{(\overline{{M}_{\rm C}})^{1/2} \, \log(0.4N)},
\end{equation}
where $N$ is the number of cluster members, $\overline{{M}_{\rm C}}$ is the mean stellar mass in units of $M_\odot$, and $R_{\rm h}$ is the half-mass radius in parsecs.

The half-mass radius is derived from the structural parameters using \citep{Sableviciute_2006}
\begin{equation}
R_{\rm h} = 0.547 \, r_{\rm c} \left(\frac{r_{\rm t}}{r_{\rm c}}\right)^{0.486},
\end{equation}
where $r_{\rm c}$ is the core radius and $r_{\rm t}$ indicate to the tidal radius given by \citep{Jeffries2001} like $r_{\rm t} = 1.46\sqrt[3]{M_C}$. Therefore, $r_{\rm t}$ is found to be 11.08 $\pm$ 3.33 pc and 7.62 $\pm$ 2.76 pc for both NGC 2423 and NGC 2482, respectively. Using the derived structural and mass parameters, we obtained half-mass radii of $R_{\rm h} = \mathrm{1.57}$~pc for NGC 2423 and $R_{\rm h} = \mathrm{3.57}$~pc for NGC 2482. The corresponding dynamical relaxation times are $T_{\rm relax} = \mathrm{18}$~Myr and $T_{\rm relax} = \mathrm{35}$~Myr, respectively. Given that these timescales are significantly shorter than the estimated ages of the clusters, both NGC 2423 and NGC 2482 can be considered dynamically relaxed systems.

\textcolor{black}{The derived relaxation times depend on the adopted observational depth and mean stellar mass. Since our Gaia DR3 membership sample is limited to $G\leq20.5$~mag, the estimated $T_{\rm relax}$ values represent the observed stellar population and may be affected by incompleteness at the low-mass end. The absence of significant mass segregation in NGC 2482 may also reflect its limited sample size and observable mass range.}

\section{Discussion}

The astrometric, photometric, and kinematic parameters derived for NGC 2423 and NGC 2482 allow both systems to be discussed within the broader context of intermediate-age open clusters. The mean parallaxes of $\varpi = 1.060\pm0.035$ mas for NGC 2423 and 0.735 $\pm$ 0.061 mas for NGC 2482 correspond to astrometric distances of 944 $\pm$ 31 pc and $1361\pm113$ pc, respectively. These values are consistent, within uncertainties, with the photometric distances obtained from isochrone fitting ($892\pm50$ pc and $1311\pm146$ pc). The adopted {\sc PARSEC} models yield ages of $1250\pm100$ Myr for NGC 2423 and $445\pm50$ Myr for NGC 2482, together with modest reddening values (Table~\ref{all_results}). Overall, these parameters fall within the range reported by recent \emph{Gaia}-based studies, with remaining differences plausibly related to membership selection, photometric depth, and model assumptions (e.g. \citealt{Cantat_2020, Tang2019, Boffin2022}).

The kinematic analysis is based on the mean proper motions and on \emph{Gaia} DR3 radial velocities available for subsets of confirmed members (Table~\ref{all_results}). Consistent with the quantitative criteria defined in Section~\ref{Introduction}, we utilize specific phase-space boundaries to establish Hyades-like membership. \textcolor{black}{Both investigated clusters have $L_z$ values within the range occupied by previously identified Hyades-like clusters. However, this angular-momentum range is not unique to the Hyades stream and can also be expected for near-circular thin-disc orbits near the
solar circle. We therefore do not use $L_z$ alone as a defining criterion for Hyades-stream membership. Instead, it is used as an action-space diagnostic to compare the two clusters with the Hyades-like population reported in the literature. NGC 2482 also shows a close correspondence in the $U$--$V$ plane, whereas NGC 2423 has a similar rotational velocity but a substantially different $U$ component. Thus, the two clusters show different degrees of kinematic similarity to the Hyades population.}

\textcolor{black}{In this study, the term Hyades-like is used to describe kinematic similarities with the Hyades population rather than a shared formation origin. The comparison considers both velocity-space and angular-momentum properties, while recognizing that the $L_z$ range alone is not unique to the Hyades stream.} NGC 2423 and NGC 2482 are therefore not interpreted as genetic counterparts of the Hyades cluster. This distinction is particularly relevant for NGC 2423, whose age ($\sim$ 1.25 Gyr) exceeds that of the Hyades ($\sim$ 600–700 Myr), effectively ruling out a common birth event. The adopted classification follows standard definitions of stellar streams and moving groups, where kinematic similarity does not imply spatial proximity or coeval formation \citep{1979ApJ...234..168E, Dehnen1998, Famaey2005}.

Despite their large present-day distances from the Hyades cluster, both NGC 2423 and NGC 2482 exhibit kinematic properties compatible with the Hyades moving group. Such spatial offsets are expected for moving-group members, which are defined by velocity-space structure rather than physical clustering \citep{Famaey2005}. Similar tidal extensions and kinematic associations related to the Hyades stream have been identified around Praesepe, Coma Berenices, and NGC~752, extending out to distances of $\sim$ 1 kpc \citep{Gaia_DR2, Cantat_2020, Jerabkova2021, Tang2019, Kroupa2022, Boffin2022}.

\begin{figure*}
    \centering
    \includegraphics[width=0.8\linewidth]{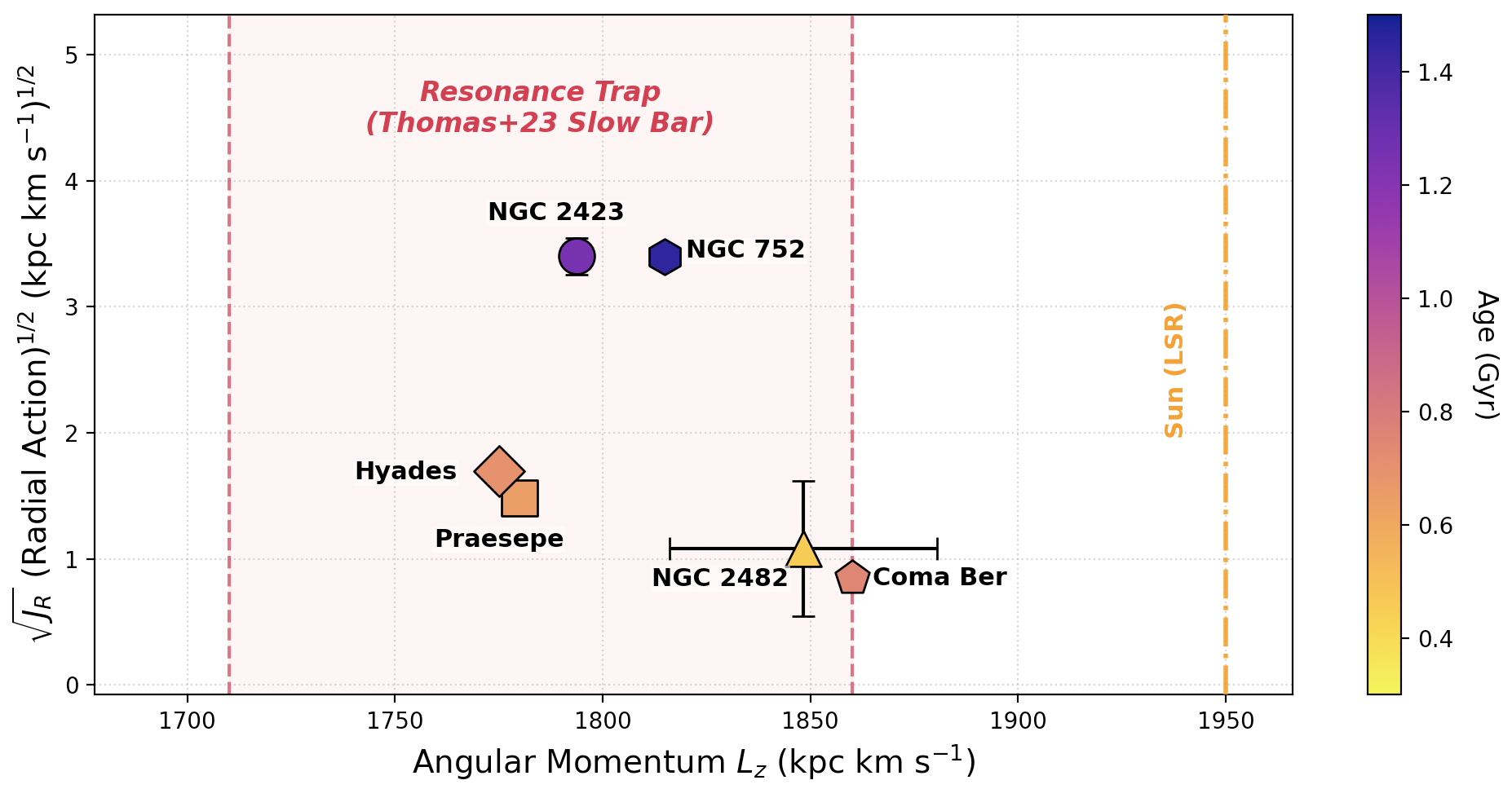}
    \caption{\textcolor{black}{Action--space diagram ($J_R$ vs. $L_z$) showing the positions of NGC 2423 and NGC 2482 relative to previously identified Hyades-like clusters. The vertical crimson dashed lines at $L_z=1710$ and $1860$~kpc~km~s$^{-1}$ indicate the literature reference interval associated with the 2:1 inner Lindblad resonance of a slowly rotating Galactic bar ($\Omega_b=39$~km~s$^{-1}$~kpc$^{-1}$; \citet{Thomas2023}). The $L_z$ interval is shown for comparison and is not used as a unique criterion for Hyades-stream membership. Colours denote cluster ages, and the Sun is marked at $L_z\approx1950$~kpc~km~s$^{-1}$.}}
    \label{fig:resonance}
\end{figure*}

To place this behavior in a broader dynamical context, the vertical angular momentum $L_z$ and radial action $J_R$ were estimated using the epicyclic approximation appropriate for near-circular disk orbits \citep{binney08}. Figure~\ref{fig:resonance} shows the distribution of NGC 2423 and NGC 2482 in action space ($\sqrt{J_R}$ versus $L_z$), together with reference clusters associated with the Hyades stream, including Praesepe, Coma Berenices, and NGC~752 \citep{Boffin2022}. \textcolor{black}{To place the kinematic properties of the two clusters in a broader action-space context, we estimated the vertical angular momentum $L_z$ and radial action $J_R$ using the epicyclic approximation appropriate for near-circular disc orbits. Figure~\ref{fig:resonance} compares the locations of NGC 2423 and NGC 2482 with those of previously studied Hyades-like clusters. Both target clusters occupy the same general $L_z$ range as the reference Hyades-like population. However, this range is also expected for a broad population of thin-disc clusters at Galactocentric radii close to the solar circle. Consequently, the $L_z$ overlap should be regarded as a consistency check rather than as a distinctive signature of Hyades-stream membership.}

\textcolor{black}{The present-day action-space locations of NGC 2423 and NGC 2482 overlap with the narrow $L_z$ interval occupied by several Hyades-like clusters. This interval has been associated in previous studies with the 2:1 inner Lindblad resonance of a slowly rotating Galactic bar \citep{Boffin2022, Thomas2023}. In Figure~\ref{fig:resonance}, we therefore use the resonance-related interval reported by \citet{Thomas2023} as a reference for comparison with the locations of our clusters. This comparison is based solely on their present-day action-space coordinates and should not be interpreted as a dynamical demonstration of resonance trapping or bar-driven evolution.}

\textcolor{black}{Our orbit integrations were performed using MWPotential2014, a static and axisymmetric Galactic potential that does not include a bar or spiral structure. These integrations are therefore useful for characterizing the orbital properties of the clusters, but they cannot directly test the interaction between the clusters and a Galactic bar or establish resonance capture. The consistency of the cluster locations with the literature resonance interval may support a possible dynamical interpretation of their Hyades-like kinematics, but a direct demonstration of a bar-resonance origin would require orbit modelling in a time-dependent, non-axisymmetric Galactic potential. We therefore regard the bar-resonance interpretation as a plausible association rather than a demonstrated causal origin.}

From a broader perspective, the relevance of NGC 2423 and NGC 2482 lies in their role as well-characterized open clusters that occupy a sparsely sampled region of age–action space within the Hyades moving group. Their distinct ages, combined with consistent kinematic properties, make them valuable reference points for examining how resonance-related velocity structures persist across clusters with different evolutionary states. In this sense, the two systems do not introduce new dynamical phenomena, but rather provide additional, independent constraints supporting the interpretation of the Hyades stream as a long-lived dynamical feature of the Galactic disk.

\section{Summary and Conclusion}

We present a homogeneous photometric, astrometric, and kinematic analysis of the two Hyades-like open clusters NGC 2423 and NGC 2482 based on data from {\it Gaia} DR3. By combining high-precision astrometry with photometry, we derive refined cluster memberships and investigate their fundamental parameters and stellar content in a consistent framework. Particular emphasis is placed on assessing their Hyades-like nature through comparisons of their spatial, kinematic, and evolutionary properties. The main results of this study can be summarized as follows.
\begin{itemize}
    \item From Gaussian fits to the astrometric distributions of the selected members, the mean proper motions and parallaxes were derived as $\mu_{\alpha}\cos\delta=-0.750\pm0.134$ mas yr$^{-1}$, $\mu_{\delta} = -3.578\pm0.143$ mas yr$^{-1}$, and $\varpi=1.060\pm0.035$ mas for NGC 2423, and $\mu_{\alpha}\cos\delta=-4.669\pm0.070$ mas yr$^{-1}$, $\mu_{\delta} = 2.250\pm0.065$ mas yr$^{-1}$, and $\varpi=0.735\pm0.061$ mas for NGC 2482.

    \item The corresponding astrometric distances are $944\pm31$ pc (NGC 2423) and $1361\pm113$ pc (NGC 2482). Isochrone fitting to the decontaminated CMDs yields $(m-M)_{\rm obs} = 9.90\pm0.10$ mag and $E(G_{\rm BP}-G_{\rm RP}) = 0.07\pm0.01$ mag for NGC 2423, and $(m-M)_{\rm obs} = 10.80\pm0.20$ mag and $E(G_{\rm BP}-G_{\rm RP}) = 0.10\pm0.02$ mag for NGC 2482. The implied $E(B-V)$ values are 0.054 and 0.078, with photometric distances of $892\pm50$ pc and $1311\pm146$ pc, respectively (Table~\ref{all_results}).

    \item The best-fitting {\sc PARSEC} ages are $1250\pm100$ Myr for NGC 2423 and $445\pm50$ Myr for NGC 2482. These values are broadly consistent with recent {\it Gaia}-based studies, with expected scatter among literature estimates arising from differences in adopted membership samples, photometric depth, and model assumptions (Table~\ref{tab:literature}).

    \item Mean radial velocities based on {\it Gaia} DR3 measurements of cluster members are $\langle V_{\rm r}\rangle=18.54\pm0.08$ km s$^{-1}$ for NGC 2423 (188 stars) and $\langle V_{\rm r}\rangle=39.04\pm0.14$ km s$^{-1}$ for NGC 2482 (29 stars). The more limited radial-velocity sample for NGC 2482 should be considered when comparing dispersion-related quantities.
    
    \item Mass segregation analyses reveal different states for the two clusters: NGC 2423 shows clear signatures with high-mass stars ($M > 1\,M_{\odot}$) being more centrally concentrated, while NGC 2482 shows no significant difference in the distributions of low and high-mass stars. The mass function slopes are $\alpha=2.41\pm0.06$ and $\alpha=2.22\pm0.03$, respectively. The condition $t_{\rm age} \gg t_{\rm relax}$ confirms that the segregation in NGC 2423 is of dynamical origin.

    \item Backward orbit integrations over the cluster ages suggest that NGC 2423 and NGC 2482 formed at Galactocentric radii close to their present-day locations, following low-eccentricity disc-like orbits ($e<0.1$) and remaining confined near the Galactic plane ($|Z_{\rm max}|<0.2$~kpc), consistent with dynamically regular thin-disc populations.

    \item The estimated dynamical relaxation times of NGC 2423 and NGC 2482, $T_{\rm relax} = 18$~Myr and $T_{\rm relax} = 35$~Myr respectively, are significantly shorter than their corresponding cluster ages, indicating that both systems have reached a dynamically relaxed state and no longer retain their initial dynamical configurations. The derived half-mass radii are $R_{\rm h} = 1.57$~pc for NGC 2423 and $R_{\rm h} = 3.57$~pc for NGC 2482.

    \item \textcolor{black}{In action space, NGC 2423 and NGC 2482 occupy the same general $L_z$ range as previously identified Hyades-like clusters. However, this angular-momentum range is not unique to the Hyades stream for clusters on near-circular orbits near the solar circle. NGC 2482 provides the stronger velocity-space correspondence with the Hyades population, while NGC 2423 shares its characteristic rotational velocity and angular momentum but is offset in $U$. Thus, the action-space similarity provides supporting context for a Hyades-like interpretation, rather than a standalone criterion for stream membership.}
\end{itemize}

Overall, the derived parameters provide a consistent set of distances, reddening values, ages, and mean kinematics for NGC 2423 and NGC 2482 from a uniform {\it Gaia} DR3-based analysis. Improved radial-velocity completeness and deeper photometry, particularly at the low-mass end and in the cluster outskirts, would further reduce uncertainties in the inferred structural and kinematic trends.

\section*{Acknowledgements}
We thank the anonymous referee for the valuable and constructive feedback, which has helped us improve the manuscript. This study presents results derived from the European Space Agency (ESA) space mission Gaia. The data from $Gaia$ are processed by the $Gaia$ Data Processing and Analysis Consortium (DPAC). Financial support for DPAC is provided by national institutions, primarily those participating in the $Gaia$ Multi-Lateral Agreement (MLA). For additional information, the official $Gaia$ mission website can be accessed at \url{https://www.cosmos.esa.int/gaia}, and the $Gaia$ archive is available at \url{https://archives.esac.esa.int/gaia}.

\appendix

\section{Spatial and Kinematic Distribution of Member Stars as a Function of Parallax}
\label{app:parallax_tail}

\begin{figure*}[!ht]
\centering
\includegraphics[width=0.85\linewidth]{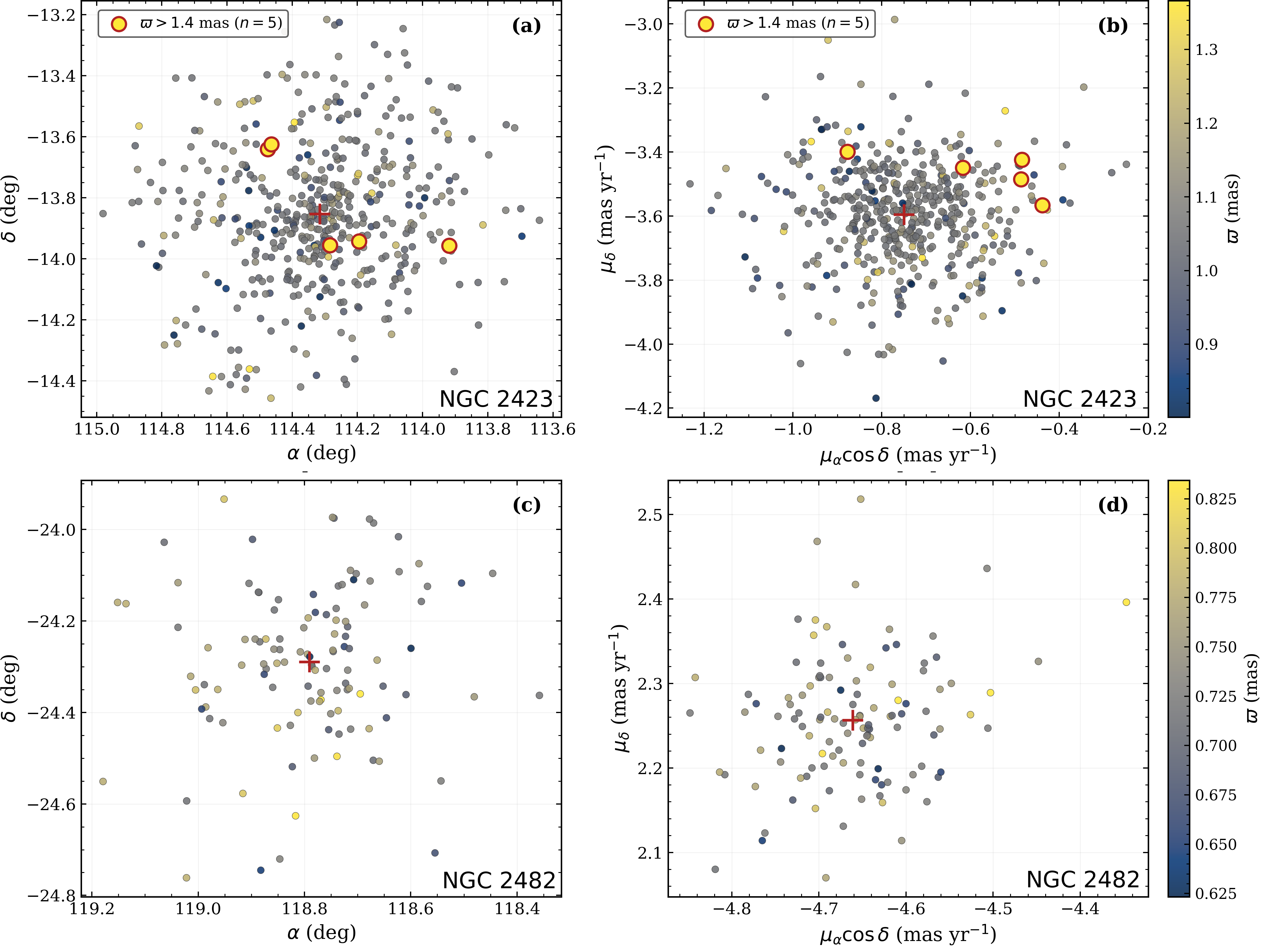}
\caption{Spatial ($\alpha$, $\delta$; left) and proper-motion ($\mu_\alpha\cos\delta$, $\mu_\delta$; right) distributions of probable members of NGC 2423 (upper panels) and NGC 2482 (lower panels), colour-coded by Gaia~DR3 trigonometric parallax. Red crosses indicate the mean position and proper motion of each cluster. The five NGC 2423 stars with $\varpi > 1.4$~mas are marked with red-outlined symbols and are likely residual field-star contaminants.}

\label{fig:appendix_parallax}
\end{figure*}

\textcolor{black}{As noted in Section~5 and by the referee, the parallax distribution of NGC 2423 members (Figure~\ref{fig: Plx}, upper panel) shows a modest excess of stars beyond $\varpi \sim 1.4$~mas, extending beyond the core of the fitted Gaussian. To determine whether these high-parallax sources are genuine cluster members or foreground contaminants retained by the $P \ge 0.5$ membership threshold, we examined the spatial and proper-motion distributions of the probable members of both clusters, colour-coded by individual trigonometric parallax, as shown in Figure~\ref{fig:appendix_parallax}.}

\textcolor{black}{For NGC 2423 (upper panels), the five stars with $\varpi > 1.4$~mas, highlighted with red-outlined markers, are spatially distributed across the field and show no preferential concentration toward the cluster centre, unlike the bulk of the membership sample. Their proper motions are also dispersed over a wider range ($\mu_\alpha\cos\delta \approx -0.9$ to $-0.4$~mas yr$^{-1}$) than the tight kinematic distribution of the remaining 483 members. These stars do not form a coherent spatial or kinematic subgroup. Their spatial and kinematic distributions therefore suggest that they are residual field-star contaminants admitted by the $P \ge 0.5$ membership threshold rather than a distinct, closer population of genuine cluster members. Since these five stars represent only $\sim 1\%$ of the adopted NGC 2423 membership (5 out of 488 stars), their effect on the derived cluster parameters is expected to be small. We further verified that excluding them changes the mean parallax by less than $0.005$~mas, which is well within the quoted uncertainty.}

\textcolor{black}{For NGC 2482 (lower panels), no comparable high-parallax excess is observed. All 116 member stars occupy a narrow and unimodal parallax distribution consistent with the Gaussian fit shown in Figure~\ref{fig: Plx} (lower panel), with no sources showing a similar excess. The spatial and proper-motion distributions are also compact, with no clear outliers separated from the bulk of the membership sample. This behaviour is consistent with the more homogeneous membership distribution of NGC 2482 and the sharper $P$-distribution shown in Figure~\ref{fig:prob_hist}.}

\bibliography{references}

\end{document}